%% file: mainRRC.tex
\documentclass{article}
\usepackage{arxiv}
\usepackage{bbm,fullpage}
\usepackage[T1]{fontenc}
\usepackage[utf8]{inputenc}
\usepackage[english]{babel}
\usepackage{amsmath,amssymb,amsfonts,amsthm}
\usepackage{cmap}
\usepackage{microtype}
\usepackage{mathtools}
\usepackage{mathrsfs}
\usepackage{bm}

\usepackage{url}            
\usepackage{booktabs}       
\usepackage{nicefrac}       
\usepackage{microtype}      

\usepackage[linesnumbered,ruled,vlined]{algorithm2e}
\SetKwInput{KwData}{Input}
\SetKwInput{KwResult}{Output}
\SetKwInput{KwAssumption}{Assumption}
\usepackage{xcolor}
\usepackage[colorlinks=true]{hyperref}
\usepackage{graphicx}
\usepackage{tabularx}
\usepackage{booktabs}

\usepackage[inline,shortlabels]{enumitem}

\usepackage[linesnumbered,ruled,vlined]{algorithm2e}

\SetCommentSty{mycommfont}
\SetKwInput{KwData}{Input}
\SetKwInput{KwResult}{Output}
\SetKwInput{KwAssumption}{Assumption}

\newtheorem{theorem}{Theorem}

\makeatletter
\newtheorem*{rep@theorem}{\rep@title}
\newcommand{\newreptheorem}[2]{%
\newenvironment{rep#1}[1]{%
 \def\rep@title{#2 \ref{##1}}%
 \begin{rep@theorem}}%
 {\end{rep@theorem}}}
\makeatother

\newreptheorem{theorem}{Theorem}

\newtheorem{proposition}{Proposition}
\newtheorem{problem}[proposition]{Problem}
\newtheorem{lemma}[proposition]{Lemma}
\newtheorem{corollary}[proposition]{Corollary}

\newtheorem{definition}[proposition]{Definition}
\newtheorem{remark}[proposition]{Remark}

\newtheorem{assumption}{Assumption}

\makeatletter
\newtheorem*{rep@assumption}{\rep@title}
\newcommand{\newrepassumption}[2]{%
\newenvironment{rep#1}[1]{%
 \def\rep@title{#2 \ref{##1}}%
 \begin{rep@assumption}}%
 {\end{rep@assumption}}}
\makeatother

\newrepassumption{assumption}{Assumption}

\newcolumntype{Y}{>{\centering\arraybackslash}X}

\input{macro}

\title{Solving parametric systems of polynomial equations over the
reals through Hermite matrices}

\author{Huu Phuoc {Le} \\
 Sorbonne Universit\'e, \textsc{CNRS},\\
    Laboratoire d'Informatique de Paris~6, \textsc{LIP6}, \\
    \'Equipe \textsc{PolSys} \\
    F-75252, Paris Cedex 05, France \\
    \texttt{huu-phuoc.le@lip6.fr}
\And
Mohab {Safey El Din} \\
Sorbonne Universit\'e, \textsc{CNRS},\\
Laboratoire d'Informatique de Paris~6, \textsc{LIP6}, \\
\'Equipe \textsc{PolSys} \\
F-75252, Paris Cedex 05, France \\
\texttt{mohab.safey@lip6.fr}}

\date{\today}

\begin{document}
\maketitle

\begin{abstract}
We design a new algorithm for solving parametric systems of equations
having finitely many complex solutions for generic values of the
parameters. More precisely, let $\ff = (f_1, \ldots, f_m)\subset
\QQ[\by][\bx]$ with $\by = (y_1, \ldots, y_{t})$ and $\bx = (x_1,
\ldots, x_{n})$, $\cV\subset \CC^t \times \CC^n$ be the algebraic set
defined by the simultaneous vanishing of the $f_i$'s and $\pi$ be the
projection $(\by, \bx) \to \by$. Under the assumptions that $\ff$
admits finitely many complex solutions when specializing $\by$ to
generic values and that the ideal generated by $\ff$ is radical, we
solve the following algorithmic problem. On input $\ff$, we compute
{\em semi-algebraic formulas} defining open semi-algebraic sets
$\mathcal{S}_1, \ldots, \mathcal{S}_\ell$ in the parameters' space
$\RR^t$ such that $\cup_{i=1}^\ell \mathcal{S}_i$ is dense in $\RR^t$
and, for $1\leq i \leq \ell$, the number of real points in $\cV\cap
\pi^{-1}(\bma)$ is invariant when $\bma$ ranges over $\mathcal{S}_i$.

This algorithm exploits special properties of some well chosen
monomial bases in the quotient algebra $\QQ(\by)[\bx] / I$ where
$I\subset \QQ(\by)[\bx]$ is the ideal generated by $\ff$ in
$\QQ(\by)[\bx]$ as well as the specialization property of the
so-called Hermite matrices which represent Hermite's quadratic
forms. This allows us to obtain ``compact'' representations of the
semi-algebraic sets $\mathcal{S}_i$ by means of semi-algebraic
formulas encoding the signature of a given symmetric matrix.
 
When $\ff$ satisfies extra genericity assumptions (such as regularity),
we use the theory of Gr\"obner bases to derive complexity
bounds both on the number of arithmetic operations in $\QQ$ and the
degree of the output polynomials. More precisely, letting 
$d$ be the maximal degrees of the $f_i$'s and $\mathfrak{D} = n(d-1)d^n$, 
we prove that, on a generic input
$\ff=(f_1,\ldots,f_n)$, one can compute those semi-algebraic formulas using $O\
{\widetilde{~}}\left ( \binom{t+\mathfrak{D}}{t}\ 2^{3t}\ n^{2t+1}
d^{3nt+2(n+t)+1} \right )$ arithmetic operations in $\QQ$ and that the
polynomials involved in these formulas have degree bounded by
$\mathfrak{D}$.

We report on practical experiments which illustrate the efficiency of
this algorithm, both on generic parametric systems and parametric
systems coming from applications since it allows us to solve systems
which were out of reach on the current state-of-the-art.
\end{abstract}


\footnotesize{\thanks{Mohab Safey El Din and Huu Phuoc Le are supported by the
ANR grants ANR-18-CE33-0011 \textsc{Sesame}, and ANR-19-CE40-0018
\textsc{De Rerum Natura}, the joint ANR-FWF ANR-19-CE48-0015
\textsc{ECARP} project, the PGMO grant \textsc{CAMiSAdo}, the
European Union’s Horizon 2020 research and innovative training network
program under the Marie Skłodowska-Curie grant agreement N° 813211
(POEMA) and the Grant FA8665-20-1-7029 of the EOARD-AFOSR.}}

\input{intro}

\input{preliminaries}

\input{samplepoints}

\input{hermite}

\input{algorithm}

\input{complexity}

\input{experiments}

\paragraph{Acknowledgments} 
We thank the anonymous reviewers for their comments which helped to
improve a lot the initial submission.

\bibliography{mybib}
\bibliographystyle{acm}
\end{document}

%% file: macro.tex
\newcommand{\CC}{\mathbb{C}}
\newcommand{\ZZ}{\mathbb{Z}}
\newcommand{\FF}{{\mathbb{F}}}
\newcommand{\RR}{\mathbb{R}}
\newcommand{\QQ}{\mathbb{Q}}
\newcommand{\KK}{\mathbb{K}}

\newcommand{\VV}{V}

\newcommand{\sign}{\mathrm{sign}\;}

\newcommand{\sing}{\mathrm{sing}}

\newcommand{\crit}{\mathrm{crit}}

\newcommand{\todo}[1]{{{#1}}}
\newcommand{\revi}[1]{{{#1}}}

\newcommand{\bx}{{\bm{x}}}

\newcommand{\by}{{\bm{y}}}

\newcommand{\ff}{{\bm{f}}}

\newcommand{\bma}{{\eta}}

\newcommand{\DRL}{\mathrm{grevlex}}
\newcommand{\LEX}{{lex}}

\newcommand{\cL}{{\mathcal{L}}}

\newcommand{\cC}{{\mathcal{C}}}

\newcommand{\cG}{{\mathcal{G}}}
\newcommand{\cH}{{\mathcal{H}}}

\newcommand{\cS}{{\mathcal{S}}}

\newcommand{\cV}{{\mathcal{V}}}
\newcommand{\cW}{{\mathcal{W}}}

\newcommand{\nf}{{w}}

\newcommand{\x}{{\bm{x}}}

\newcommand{\NF}{\mathrm{NF}}

\newcommand{\lm}{\mathrm{lm}}
\newcommand{\lc}{\mathrm{lc}}
\newcommand{\trace}{\mathrm{trace}}

%% file: intro.tex
\section{Introduction}
\label{section:intro}
\subsection{Problem statement and motivations}
\label{ssection:statement}

In the whole paper, $\QQ$, $\RR$ and $\CC$ denote respectively the
fields of rational, real and complex numbers.

Let $\ff=(f_1,\ldots,f_m)$ be a polynomial sequence in $\QQ[\by][\bx]$
where the indeterminates $\by=(y_1,\ldots,y_t)$ are considered as {\em
parameters} and $\bx=(x_1,\ldots,x_n)$ are considered as {\em
variables}. We denote by $\cV \subset \CC^{t}\times \CC^n$ the
(complex) algebraic set defined by $f_1=\cdots=f_m=0$ and by
$\cV_{\RR}$ its real trace $\cV\cap \RR^{t+n}$. We consider also the
projection on the parameter space $\by$
\[
  \pi:
  \begin{array}{rl}
    \CC^t \times \CC^n &\to \CC^t, \\
    (\by,\bx) & \mapsto \by.
  \end{array}
\]
Further, we say that $\ff$ satisfies Assumption \eqref{assumption:A}
when the following holds.
\begin{assumption}
  \label{assumption:A}
  There exists a non-empty Zariski open subset $\mathcal{O} \subset
\CC^t$ such that $\pi^{-1}(\bma)\cap\cV$ is non-empty and finite for
any $\bma \in \mathcal{O}$.
\end{assumption}
In other words, assuming \eqref{assumption:A} ensures that, for a
generic value $\bma$ of the parameters, the sequence $\ff(\bma,\cdot)$
defines a finite algebraic set and hence finitely many real points.
Note that, it is easy to prove that one can choose $\mathcal{O}$ in a
way that the number of complex solutions to the entries of
$\ff(\bma,\cdot)$ is invariant when $\bma$ ranges over $\mathcal{O}$
(e.g. using the theory of Gr\"obner basis). This is no more the case
when considering real solutions whose number may vary when $\bma$
ranges over $\mathcal{O}$.

By Hardt's triviality theorem \cite{Hardt80}, there exists a real
algebraic \emph{proper} subset $\mathcal{R}$ of $\RR^t$ such that, for
any non-empty connected open set $\mathcal{U}$ of $\RR^t\setminus
\mathcal{R}$ and $\bma \in \mathcal{U}$, $\pi^{-1}(\bma) \times
\mathcal{U}$ is homeomorphic with $\pi^{-1}(\mathcal{U})$.

This leads us to consider the following real root classification
problem.

\begin{problem}[Real root classification]
\label{problem:rrc}
On input $\ff$ satisfying Assumption~\eqref{assumption:A}, compute
{\em semi-algebraic} formulas (i.e. finitely many disjunctions of
conjunctions of polynomial inequalities) defining semi-algebraic sets
$\mathcal{S}_1, \ldots, \mathcal{S}_\ell$ such that
\begin{itemize}
\item[(i)] The number of real points in $\mathcal{V}\cap
\pi^{-1}(\bma)$ is invariant when $\bma$ ranges over $\mathcal{S}_i$,
for $1\leq i \leq \ell$;
\item[(ii)] The union of the $\mathcal{S}_i$'s is dense in $\RR^t$;
\end{itemize}
as well as at least one sample point $\bma_i$ in each $\mathcal{S}_i$
and the corresponding number of real points in $\cV\cap
\pi^{-1}(\bma_i)$.

A collection of semi-algebraic formulas sets is said to solve Problem
\eqref{problem:rrc} for the input $\ff$ if it defines a collection of
semi-algebraic sets $\cS_i$ satisfies the above properties (i) and
(ii).

Our output will have the form $\{(\Phi_i,\bma_i,r_i)\; | \; 1 \leq i
\leq \ell\}$ where $\Phi_i$ is a semi-algebraic formula defining the
set $\cS_i$, $\bma_i \in \QQ^t$ is a sample point of $\cS_i$ and $r_i$
is the corresponding number of real roots.
\end{problem}

A weak version of Problem~\eqref{problem:rrc} would be to compute only
a set $\{\bma_1, \ldots, \bma_{\ell}\}$ of sample points for a collection
of semi-algebraic sets $\cS_i$ solving Problem \eqref{problem:rrc} and
their corresponding numbers of real points in
$\cV\cap \pi^{-1}(\bma_j)$.


Problem~\eqref{problem:rrc} appears in many areas of engineering
sciences such as robotics or medical imagery (see, e.g.,
\cite{Yang00,Co02,YaZhe05,FMRSa08,BFJSV16}). 

In this paper, we design a new algorithm whose arithmetic complexity improves
the previously known bounds and reports on practical experiments showing that
its practical behaviour outperforms the current software state-of-the-art.

Before going further with a description of the prior works and our
contributions, we introduce the complexity model which we use. We measure only
the arithmetic complexity of algorithms, i.e., the number of arithmetic
operations $+,-,\times,\div$, in the base field $\QQ$, hence, without
  taking into account the cost of real root isolation. We use the Landau
notation:
\begin{itemize}
\item Let $f: \RR_+^{\ell} \mapsto \RR_+$ be a positive function. We
let $O(f)$ denote the class of functions $g: \RR_+^{\ell} \to \RR_+$
such that there exist $C,K \in \RR_+$ such that for all $\|x\| \ge
K$, $g(x) \leq C f(x)$, where $\|\cdot \|$ is a norm of
$\RR^{\ell}$.
\item The notation $O\ {\widetilde{~}}$ denotes the class of functions
$g:\RR_+^{\ell} \to \RR_+$ such that $ g \in O(f\log^{\kappa}(f))$ for
some $\kappa > 0 $.
\end{itemize}
Further, the notation $\omega$ always stands for the exponent constant
of the matrix multiplication, i.e., the smallest positive number such
that the product of two matrices in $\QQ^{N\times N}$ can be done
using $O\left (N^{\omega}\right)$ arithmetic operations in
$\QQ$. The value of $\omega$ can be bounded from above by
$2.37286$, which is established in \cite{Alman21}.




\subsection{Prior works}
A first approach to Problem~\eqref{problem:rrc} would be to compute a
cylindrical algebraic decomposition (CAD) of $\RR^t\times \RR^n$
adapted to $\ff$ using e.g. Collins' algorithm (and its more recent
improvements) ; see~\cite{Collins76}. While, up to our knowledge,
there is no clear reference for this fact, the cylindrical structure
of the cells of the CAD will imply that their projection on the
parameters' space $\RR^t$ define semi-algebraic sets enjoying the
properties needed to solve Problem~\eqref{problem:rrc}. However, the
doubly exponential complexity of CAD both in terms of runtime and
output size \cite{Dav88, Dav07} makes it difficult to use in
practice.

A more popular approach consists in computing polynomials $h_1,
\ldots, h_r$ in $\QQ[\by]$ such that $\cup_{i=1}^r V(h_i)\cap \RR^t$
contains the boundaries of semi-algebraic sets $\mathcal{S}_1, \ldots,
\mathcal{S}_\ell$ enjoying the properties required to solve
Problem~\eqref{problem:rrc}. Next, one needs to compute semi-algebraic
descriptions of the connected components of $\RR^t \setminus
\cup_{i=1}^r V(h_i)$ as well as sample points in these connected
components.  This is basically the approach followed by
\cite{YangXia05} (the $h_i$'s are called {\em border polynomials}) and
\cite{LaRo07} (the set $\cup_{i=1}^r V(h_i)$ is called {\em
discriminant variety}) under the assumption that $\langle \ff \rangle$
is a radical ideal. Note that both \cite{YangXia05} and
\cite{LaRo07} provide algorithms that can handle variants of
Problem~\eqref{problem:rrc} allowing inequalities. In this paper, we
focus on the situation where we only have equations in our input
parametric system.

When $\langle \ff \rangle$ is radical and the restriction of $\pi$ to $\cV\cap
\RR^t\times\RR^n$ is proper, one can easily prove using a semi-algebraic version
of Thom's isotopy lemma \cite{CoSh92} that one can choose $\cup_{i=1}^r V(h_i)$
to be the set of critical values of the restriction of $\pi$ to $\cV$ (see e.g.
\cite{BFJSV16}). If $\ff$ is a regular sequence (hence $m = n$), the critical
set of the restriction of $\pi$ to $\cV$ is defined as the intersection of $\cV$
with the hypersurface defined by the vanishing of the determinant of the
Jacobian matrix of $\ff$ with respect to the variables $\bx$. When $d$ dominates
the degrees of the entries of $\ff$, B\'ezout's theorem allows us to state that
the degree of this set is bounded above by $n(d-1)d^n$.

It is worth noticing that, usually, this approach is used only to
solve the aforementioned {\em weak} version of
Problem~\eqref{problem:rrc} as getting a semi-algebraic description of
the connected components of $\RR^t \setminus \cup_{i=1}^r V(h_i)$
through CAD is too expensive when $t\geq 4$ (still, because of the
doubly exponential complexity of CAD). Under the above assumptions and
notation, the output degree of the polynomials in such formulas would
be bounded by $\left (n(d-1)d^n\right )^{2^{O(t)}}$.

An alternative would be to use {\em parametric} roadmap algorithms to
do such computations using e.g. \cite[Chap. 16]{BPR} to compute
semi-algebraic representations of the connected components of $\RR^t
\setminus \cup_{i=1}^r V(h_i)$.  Under the above extra assumptions,
this would result in output formulas involving polynomials of degree
bounded by $\left (n(d-1)d^n\right )^{O(t^3)}$ using $\left
(n(d-1)d^n\right )^{O(t^4)}$ arithmetic operations (see \cite[Theorem
16.13]{BPR}). Note that the output degrees are by several orders of
magnitude larger than $n(d-1)d^n$ which bounds the degree of the set
of critical values of the restriction of $\pi$ to $\cV$.

Hence, one topical algorithmic issue is to design an efficient algorithm for
solving Problem~\eqref{problem:rrc} which would output semi-algebraic formulas
of degree bounded by $n(d-1)d^n$ (using a number of arithmetic operations
polynomial in this quantity). At this stage of our exposition, this is not clear
that it is doable. \todo{Actually, admittedly ``folklore'' algorithms in symbolic
computation already allow one to achieve such a result.}

\revi{Using the (probabilistic) algorithm of \cite{Sc03}, one can compute a
  rational parametrization of $\cV = V(\ff)$ with respect to the
  $\bx$-variables, i.e. a sequence of polynomials $(\nf, v_1, \ldots, v_n)$ in
  $\QQ(\by)[u]$ where $u$ is a new variable, such that the constructible set
  $\mathcal{Z} \subset \CC^t\times \CC^n$ of every point
\[\left (\bma,\frac{v_1}{\partial \nf / \partial
      u}(\bma, \vartheta), \ldots, \frac{v_n}{\partial \nf / \partial u}(\bma,
    \vartheta)\right ),\] where $(\bma,\vartheta) \in \CC^t \times \CC$ such
that $w(\bma,\vartheta) = 0$ and $\bma$ does not cancel $\partial w/ \partial u$
and any denominator of $(\nf, v_1,\ldots,v_n)$, is Zariski dense in $\cV$, i.e.,
the Zariski closure of $\mathcal{Z}$ coincides with $\cV$.

{The bi-rational equivalence between $\mathcal{Z}$ and its projection on the $(u,
\by)$-space implies that semi-algebraic formulas solving
Problem~\eqref{problem:rrc} can be obtained through the computation of the
subresultant sequence associated to $\left (\nf, \frac{\partial \nf}{\partial
    u}\right )$ (see e.g.~\cite[Chap. 4]{BPR}). 
Combining the complexity results of~\cite{Sc03} to compute a rational
parametrization of $\cV$ with those of~\cite[Chap. 4]{BPR} for computing
subresultants we obtain that this algorithm uses  
\[O\ {\widetilde{~}}\left (\binom{t+2d^{2n}}{t}\ 2^{5t} \ d^{5nt+3n} \right)\]
arithmetic operations in $\QQ$, and that the semi-algebraic formulas computed by
this algorithm involve polynomials in $\QQ[\by]$ of degree bounded by $2d^{2n}$.
Recall that the degree of the critical locus of the restriction of $\pi$ to
$\cV$ is bounded by $n(d-1)d^n$. Hence, computing semi-algebraic formulas
solving Problem~\eqref{problem:rrc} involving polynomials of degrees in $O(d^n)$
through an efficient algorithm reflecting this complexity gain is still an open
problem.}
}

\subsection{Main results}


Basically, our main result is to provide a new algorithm solving
Problem~\eqref{problem:rrc} when $\langle \ff \rangle$ is radical and
assumption~\eqref{assumption:A} holds. Under some genericity assumptions, we
prove that it outputs formulas involing polynomials of degree in $O(d^n)$ with a
better arithmetic complexity than what was previously known.

\begin{theorem}
  \label{thm:main}
Let $\CC[\bx,\by]_d$ be the set of polynomials in $\CC[\bx,\by]$
having total degree bounded by $d$ and set $\mathfrak{D} = n(d-1)d^n$.

There exists a non-empty Zariski open set $\mathscr{F}\subset
\CC[\bx,\by]_d^n$ such that for $\ff=(f_1,\ldots,f_n) \in
\mathscr{F} \cap \QQ[\bx,\by]^n$, the following holds:
\begin{itemize}
\item[i)] There exists an algorithm that computes a solution for the
  weak-version of Problem \eqref{problem:rrc} within
    \[O\ {\widetilde{~}}\left (\binom{t+\mathfrak{D}}{t} \
   2^{3t}\ n^{2t+1} d^{2nt+n+2t+1} \right ).\]
  arithmetic operations in $\QQ$.
\item[ii)] There exists a probabilistic algorithm that returns
  the formulas of a collection of semi-algebraic sets solving
  Problem \eqref{problem:rrc} within
\[O\ {\widetilde{~}}\left ( \binom{t+\mathfrak{D}}{t}\ 2^{3t}\
n^{2t+1} d^{3nt+2(n+t)+1} \right )\]
arithmetic operations in $\QQ$ in case of success.
\item[iii)] The semi-algebraic descriptions output by the above
algorithm involves polynomials in $\QQ[\by]$ of degree bounded by
$\mathfrak{D}$.
\end{itemize}
\end{theorem}
We note that the binomial coefficient $\binom{t+\mathfrak{D}}{t}$ is
bounded from above by $\mathfrak{D}^t \simeq n^td^{nt+t}$. Therefore,
the complexities given in the items i) and ii) of
Theorem~\ref{thm:main} can be bounded by $O\ {\widetilde{~}}\left (
2^{3t}\ n^{3t} d^{3nt} \right )$ and $O\
{\widetilde{~}}\left ( 2^{3t}\ n^{3t} d^{4nt} \right )$ respectively.

\todo{We also implemented this algorithm to illustrate its practical behaviour
  and compare it with the state-of-the-art software within the {\sc Maple}
  packages {\sc RootFinding[Parametric]} and {\sc
    RegularChains[ParametricSystemTools]}. We report on experiments showing that
  our implementation outperforms these packages, which is justified by our
  complexity result.

}

\todo{The key ingredient on which one relies to obtain these results is a set of
  well-known properties of {\em Hermite quadratic forms} to count the real roots
  of zero-dimensional ideals. The use of such quadratic forms for counting the
  number of real solutions was introduced in \cite{Hermite56} and then later on
  generalized by \cite{PRS93} and used in \cite{Ro99}. We refer
  to~\cite[Theorem 4.102]{BPR} for the explicit relation between the number of
  real roots of a zero-dimensional algebraic set and the signature of these
  quadratic forms and to \cite[Algo. 8.43]{BPR} for an algorithm computing
  these signatures.}



We first slightly extend the definition of Hermite's quadratic forms
and Hermite's matrices to the context of parametric systems; we call
them parametric Hermite quadratic forms and parametric Hermite
matrices. This is easily done since the ideal of $\QQ(\by)[\bx]$
generated by $\ff$, considering $\QQ(\by)$ as the base field, has
dimension zero. We also establish natural specialization properties
for these parametric Hermite matrices.

Hence, a parametric Hermite matrix, similar to its zero-dimensional
counterpart, allows one to count respectively the number of distinct
real and complex roots at any parameters outside a strict algebraic
sets of $\RR^t$ by evaluating the signature and rank of its
specialization.

Based on this specialization property, we design two algorithms for
solving Problem~\eqref{problem:rrc} and also its weak version for the
input system $\ff$ which satisfies Assumption \eqref{assumption:A} and
generates a radical ideal.

Our algorithm for the weak version of Problem \eqref{problem:rrc}
reduces to the following main steps.

\begin{itemize}
\item[(a)] Compute a parametric Hermite matrix $\cH$ associated to $\ff
\subset \QQ[\by][\bx]$.

\item[(b)] Compute a set of sample points $\{\bma_1, \ldots,
\bma_{\ell}\}$ in the connected components of the semi-algebraic set of
$\RR^t$ defined by $\bm{w} \neq 0$ where $\bm{w}$ is derived from $\cH$.

This is done through the so-called critical point method (see e.g.
\cite[Chap. 12]{BPR} and references therein) which are adapted to
obtain practically fast algorithms following \cite{SaSc03}. We will
explain in detail this step in Section~\ref{section:sample-points}.

This algorithm takes as input $s$ polynomials of degree $D$ involving
$t$ variables and computes sample points per connected components in
the semi-algebraic set defined by the non-vanishing of these
polynomials using
 \[O\ \widetilde{~} \left(\binom{D + t}{t}
s^{t+1}2^{3t}D^{2t+1} \right) .\]

\item[(c)] Compute the number $r_i$ of real points in $\cV\cap
\pi^{-1}(\bma_i)$ for $1\leq i \leq \ell$.

This is done by simply evaluating the signature of the specialization
of $\cH$ at each $\bma_i$.
\end{itemize}

It is worth noting that, in the algorithm above, we obtain through
parametric Hermite matrices a polynomial $\bm{w}$ that plays the same
role as the discriminant varieties of \cite{LaRo07} or the border
polynomials of \cite{YangHx01}. We will see in the section reporting
experiments that our approach outperforms the other two on every
example we consider.

To return semi-algebraic formulas, our routine is basically the same except
instead of computing sample points in the set $\{w \neq 0\}$, one needs to
consider all principal minors of the matrix $\cH$ and compute sample points
outside the union of the vanishing sets of all these polynomials.






Another contribution of this paper is to make clear how to perform the step (a).
For this, we rely on the theory of Gr\"obner bases. More precisely, we use
specialization properties of Gr\"obner bases, similar to those already proven in
\cite{Kal97}. This leaves some freedom when running the algorithm: since we
rely on Gr\"obner bases, one may choose monomial orderings which are more
convenient for practical computations. In particular, the monomial basis of the
quotient ring $\QQ(\by)[\bx]/I$ where $I$ is the ideal generated by $\ff$ in
$\QQ(\by)[\bx]$ depends on the choice of the monomial ordering used for
Gr\"obner bases computations. We describe the behavior of our algorithm when
choosing the graded reverse lexicographical ordering whose interest for
practical computations is explained in \cite{BaSt87}. Further, we denote by
$\DRL(\bx)$ the graded reverse lexicographical ordering applied to the sequence
of the variables $\bx = (x_1, \ldots, x_n)$ (with $x_1\succ \cdots \succ x_n$).
Further, we also denote by $\succ_{\LEX}$ the lexicographical ordering.

We report, at the end of the paper, on the practical behavior of this
algorithm. We compare with two Maple packages {\sc
RootFinding[Parametric]} and {\sc
RegularChains[ParametricSystemTools]} which respectively implement the
algorithms of \cite{LaRo07} and \cite{YangXia05}. In particular, our
algorithm allows us to solve instances of Problem~\eqref{problem:rrc}
which were not tractable by the state-of-the-art as well as the actual
degrees of the polynomials in the output formula which are bounded by
$n(d-1)d^n$.

We actually prove such a statement under some generic assumptions. Our
main complexity result is stated below. Its proof is given in
Subsection \ref{ssection:complexity}, where the generic
assumptions in use are given explicitly.

\paragraph{Organization of the paper}
Section \ref{section:preliminary} reviews fundamental notions of
algebraic geometry and the theory of Gr\"obner bases that we use
further. Next, we present a dedicated algorithm for computing at least
one point per connected component of a semi-algebraic defined by a
list of inequations in Section~\ref{section:sample-points}.  Section
\ref{section:ParamHermite} lies the definition and some useful
properties of parametric Hermite matrices. In Section
\ref{section:algorithm}, we describe our algorithm for solving the
real root classification problem using this parametric Hermite
matrix. The complexity analysis of the algorithms mentioned above is
given in Section \ref{section:complexity}. Finally, in Section
\ref{section:experiments}, we report on the practical behavior of our
algorithms and illustrate its practical capabilities.


%% file: preliminaries.tex
\section{Preliminaries}
\label{section:preliminary}
In the first paragraph, we fix some notations on ideals and algebraic
sets and recall the definition of critical points associated to a
given polynomial map. 
Next, we give the
definitions of regular sequences, Hilbert series, Noether position and
proper maps, which are used later in Subsection
\ref{ssection:deg-generic}. The fourth paragraph recalls some basic properties of
Gr\"obner bases and quotient algebras of zero-dimensional ideals. We
refer to \cite{CLO} for an introductory study on the algorithmic
theory of Gr\"obner bases. In the last paragraphs, we recall
respectively the definitions of zero-dimensional parametrizations and
rational parametrizations which go
back to \cite{Kr82} and is widely used in computer algebra (see e.g.
\cite{GianniM87,GLS01,GiustiHMP95}) to represent finite algebraic
sets. 
\paragraph{Algebraic sets and critical points}
We consider a sub-field $\FF$ of $\CC$. Let $I$ be a polynomial ideal
of $\FF[x_1,\ldots,x_n]$, the algebraic subset of $\CC^n$ at which the
elements of $I$ vanish is denoted by $\VV(I)$. Conversely, for an
algebraic set $\cV\subset \CC^n$, we denote by $I(\cV)\subset
\CC[x_1,\ldots,x_n]$ the radical ideal associated to $\cV$. Given any
subset $\mathcal{A}$ of $\CC^n$, we denote by $\overline{\mathcal{A}}$
the Zariski closure of $\mathcal{A}$, i.e., the smallest algebraic set
containing $\mathcal{A}$.

A map $\varphi$ between two algebraic sets $\cV \subset \CC^n$ and
$\cW\subset \CC^s$ is a polynomial map if there exist
$\varphi_1,\ldots,\varphi_t \in \CC[x_1,\ldots,x_n]$ such that the
$\varphi(\bma)=(\varphi_1(\bma),\ldots,\varphi_s(\bma))$ for $\bma\in
\cV$.

An algebraic set $\cV$ is equi-dimensional of dimension $t$ if it is
the union of irreducible algebraic sets of dimension $t$. Let
$\varphi$ be a polynomial map from $\cV$ to another algebraic set
$\cW$. The morphism $\varphi$ is dominant if and only if the image of
every irreducible component $\cV'$ of $\cV$ by $\varphi$ is Zariski
dense in $\cW$, i.e. $\overline{\varphi(\cV')} = \cW$.

Let $\phi \in \CC[x_1,\ldots,x_n]$ which defines the polynomial
function
\[\phi :
  \begin{array}{rl}
  \CC^n & \to \CC, \\
  (x_1, \ldots, x_n) & \mapsto \phi(x_1,\ldots,x_n)
  \end{array}
\]
and $\cV\subset \CC^n$ be a smooth equi-dimensional algebraic set. We
denote by $\crit(\phi, \cV)$ the set of critical points of the
restriction of $\phi$ to $\cV$. If $c$ is the codimension of $\cV$
and $(f_1,\ldots,f_m)$ generates the vanishing ideal associated to
$\cV$, then $\crit(\phi, \cV)$ is the subset of $\cV$ at which the
Jacobian matrix associated to $(f_1, \ldots, f_m, \phi)$ has rank less
than or equal to $c$ (see, e.g., \cite[Subsection 3.1]{SaSc17}).

\paragraph{Regular sequences \& Hilbert series}
Let $\FF$ be a field and $(f_1,\ldots,f_m) \subset \FF[\bx]$ where
$\bx = (x_1,\ldots,x_n)$ and $m\leq n$ be a \emph{homogeneous}
polynomial sequence. We say that $(f_1,\ldots,f_m)\subset
\FF[\bx]$ is a regular sequence if for any $i\in
\{1,\ldots,m\}$, $f_i$ is not a zero-divisor in
$\FF[\bx]/\langle f_1,\ldots,f_{i-1}\rangle$.

The notion of regular sequences is the algebraic analogue of complete
intersection. In this paper, we focus particularly on the Hilbert
series of homogeneous regular sequences, which are recalled below.

Let $I \subset \FF[\bx]$ be a homogeneous ideal. We denote by
$\FF[\bx]_{r}$ the set of every homogeneous polynomial whose degree is
equal to $r$. Then $\FF[\bx]_{r}$ and $I\cap \FF[\bx]_{r}$ are two
$\FF$-vector spaces of dimensions $\dim_{\FF}(\FF[\bx]_r)$ and
$\dim_{\FF}(I\cap \FF[\bx]_r)$ respectively. The Hilbert series of $I$
is defined as
\[{\rm HS}_{I}(z) = \sum_{r=0}^{\infty}
(\dim_{\FF}(\FF[\bx]_{r})-\dim_{\FF}(I\cap \FF[\bx]_r)) \cdot z^r.\]


We now consider the affine polynomial sequences. Note that one can
define affine regular sequences by simply removing the homogeneity
assumption of $(f_1,\ldots,f_m)$ from the above definition. However,
as explained in \cite[Sec 1.7]{Bar-thesis}, many important properties
that hold for homogeneous regular sequences are no longer valid for
the affine ones. Therefore, in this paper, we use \cite[Definition
1.7.2]{Bar-thesis} of affine regular sequences, which is more
restrictive but allows us to preserve similar results as the
homogeneous case. We recall that definition below.

For $p\in \FF[x_1,\ldots,x_n]$, we denote by ${}^Hp$ the homogeneous
component of largest degree of $p$. A polynomial sequence
$(f_1,\ldots,f_m) \subset \FF[x_1,\ldots,x_n]$, not necessarily
homogeneous, is called a regular sequence if and only if
$({}^Hf_1,\ldots,{}^Hf_m)$ is a homogeneous regular sequence.

\paragraph{Noether position \& Properness}
Let $\FF$ be a field and $\ff=(f_1,\ldots,f_n)\subset
\FF[x_1,\ldots,x_{n+t}]$. The variables $(x_1,\ldots,x_n)$ are in
Noether position with respect to the ideal $\langle \ff \rangle$ if
their canonical images in the quotient algebra
$\FF[x_1,\ldots,x_{n+t}]/\langle \ff\rangle$ are algebraic integers
over $\FF[x_{n+1}, \ldots, x_{n+t}]$ and, moreover,
$\FF[x_{n+1},\ldots,x_{n+t}]\cap \langle \ff \rangle = \langle 0
\rangle$.

From a geometric point of view, Noether position is strongly related to
the notion of proper map below (see \cite{BFS14}).

Let $\cV$ be the algebraic set defined by $\ff \in
\RR[y_1,\ldots,y_t,x_1,\ldots,x_n]$. The restriction of the projection
$\pi:(\by,\bx) \mapsto \by$ to $\cV\cap \RR^{t+n}$ is said to be
proper if the inverse image of every compact subset of $\pi(\cV\cap
\RR^{t+n})$ is compact. If the variables $\bx=(x_1,\ldots,x_n)$ is in
Noether position with respect to $\langle \ff \rangle$, then the
projection $\pi: \cV\cap \RR^{t+n} \to \RR^t, \; (\by,\bx) \mapsto
\by$ is proper.

A point $\bma \in \RR^t$ is a non-proper point of the restriction of
$\pi$ to $\cV$ if and only $\pi^{-1}(\mathcal{U}) \cap \cV \cap
\RR^{t+n} $ is not compact for any compact neighborhood $\mathcal{U}$
of $\bma$ in $\RR^t$.




\paragraph{Gr\"obner bases and zero-dimensional ideals}
Let $\mathbb{F}$ be a field and $\overline{\mathbb{F}}$ be its
algebraic closure. We denote by $\mathbb{F}[\bm{x}]$ the polynomial
algebra in the variables $\bm{x}=(x_1,\ldots,x_n)$. We fix an
admissible monomial ordering $\succ$ (see Section 2.2, \cite{CLO})
over $\mathbb{F}[\bx]$. For a polynomial $p\in \mathbb{F}[\bm{x}]$,
the leading monomial of $p$ with respect to $\succ$ is denoted by
$\lm_{\succ}(p)$.

Given an ideal $I\subset\mathbb{F}[\bm{x}]$, the initial ideal of $I$
with respect to the ordering $\succ$ is the ideal $\langle
\lm_{\succ}(p) \; | \; p \in I \rangle$. A Gr\"obner basis $G$ of $I$
with respect to the ordering $\succ$ is a generating set of $I$ such
that the set of leading monomials $\{ \lm_{\succ}(g) \; | \; g\in G\}$
generates the initial ideal $\langle \lm_{\succ}(p) \; |\; p\in
I\rangle$.

For any polynomial $p \in \mathbb{F}[\bm{x}]$, the remainder of the
division of $p$ by $G$ using the monomial ordering $\succ$ is uniquely
defined. It is called the {\em normal form} of $p$ with respect to $G$
and is denoted by $\NF_{G}(p)$. A polynomial $p$ is reduced by $G$ if
$p$ coincides with its normal form in $G$. A Gr\"obner basis $G$ is
said to be reduced if, for any $g\in
G$, all terms of $g$ are reduced modulo the leading terms of $G$.

An ideal $I$ is said to be zero-dimensional if the algebraic set $V(I)
\subset \overline{\FF}^n$ is finite and non-empty. By \cite[Sec. 5.3,
Theorem 6]{CLO}, the quotient ring $\mathbb{F}[\bm{x}]/I$ is a
$\mathbb{F}$-vector space of finite dimension. The dimension of this
vector space is also called the algebraic degree of $I$; it coincides
with the number of points of $V(I)$ counted with multiplicities
\cite[Sec. 4.5]{BPR}. For any Gr\"obner basis of $I$, the set of
monomials in $\mathbb{F}[\bm{x}]$ which are irreducible by $G$ forms a
monomial basis, which we call $B$, of this vector space. For any $p\in
\mathbb{F}[\bx]$, the normal form of $p$ by $G$ can be interpreted as
the image of $p$ in $\FF[\bx]/I$ and is a linear combination of
elements of $B$ (with coefficients in $\mathbb{F}$). Therefore, the
operations in the quotient algebra $\mathbb{F}[\bm{x}]/I$ such as
vector additions or scalar multiplications can be computed explicitly
using the normal form reduction.

In this article, while working with polynomial systems depending on
parameters in $\QQ[\by][\bx]$, we frequently take $\mathbb{F}$ to be
the rational function field $\QQ(\by)$ and treat polynomials in
$\QQ[\by][\bx]$ as elements of $\QQ(\by)[\bx]$.

\paragraph{Zero-dimensional parametrizations}
A zero-dimensional parametrization $\mathscr{R}$ of coefficients in
$\QQ$ consists of $(a_1,\ldots,a_n) \in \QQ^n$ and a sequence of
polynomials $(w,v_1,\ldots,v_n) \in (\QQ[u])^{n+1}$ where $u =
\sum_{i=1}^n a_ix_i$ such that $w$ is square-free. The solution set of
$\mathscr{R}$, defined as
\[Z(\mathscr{R}) = \left\{\left(\frac{v_1(\vartheta)}{w'(\vartheta)},
\ldots, \frac{v_n(\vartheta)}{w'(\vartheta)}\right) \in \CC^n\;|\;
\vartheta\in \CC\text{ such that } w(\vartheta) = 0\right\},\]
is finite.


A finite algebraic set $\cV \in \CC^n$ is said to be represented by a
zero-dimensional parametrization $\mathscr{R}$ if and only if $\cV$
coincides with $Z(\mathscr{R})$. Note that the cardinality of $\cV$ is
the same as the degree of $w$ ; we also call it the degree of the
zero-dimensional parametrization.

Note that it is possible to retrieve a polynomial parametrization by
inverting the derivative $w'$ modulo $w$. Still, this rational
parametrization whose denominator is the derivative of $w$ is known to
be better for practical computations as it usually involves
coefficients with smaller bit size (see \cite{DS04}).

%% file: samplepoints.tex
\section{Computing sample points in semi-algebraic sets defined by the non-vanishing of polynomials}
\label{section:sample-points}

In this section, we study the following algorithmic problem. 
Given $(g_1, \ldots, g_s)$ in $\QQ[y_1, \ldots, y_t]$, compute 
at least one sample point per connected component of the semi-algebraic set 
$\mathcal{S}\subset \RR^t$ defined by 
\[
g_1\neq 0, \ldots, g_s\neq 0.
\]
Such sample points will be encoded with zero-dimensional
parametrizations
which we described in Section~\ref{section:preliminary}. 

The main result of this section which will be used in the sequel of this paper is 
the following.

\begin{theorem}
  \label{thm:sample-points}
Let $(g_1, \ldots, g_s)$ in $\QQ[y_1, \ldots, y_t]$ with $D \geq
\max_{1\leq i \leq s} \deg(g_i)$ and $\mathcal{S}\subset \RR^t$ be the
semi-algebraic set defined by
\[
g_1\neq 0, \ldots, g_s\neq 0.
\]
There exists a probabilistic algorithm which on input $(g_1, \ldots, g_s)$
outputs a finite family of zero-dimensional parametrizations $\mathscr{R}_1,
\ldots, \mathscr{R}_k$, all of them of degree bounded by $(2D)^t$,
which encode at most $\left( 2sD \right)^t $ points such that
$\cup_{i=1}^kZ(\mathscr{R}_i)$ meets every connected component of
$\mathcal{S}$ using
  \[
    O\ {\widetilde{~}}
    \left (
      \binom{D+t}{t} s^{t+1} 2^{3t}  D^{2t+1}
    \right ). 
  \]
arithmetic operations in $\QQ$. 
\end{theorem} 

The rest of this section is devoted to the proof of this theorem. 

\begin{proof}
By~\cite[Lemma~1]{FMRSa08}, there exists a non-empty Zariski open set
$\mathcal{A}\times \mathcal{E}\subset \CC^s\times \CC$ such that for
$(\bm{a} = (a_1, \ldots, a_s), e)\in \mathcal{A}\times \mathcal{E}\cap
\RR^s\times \RR$, the following holds. For $\mathcal{I} = \{i_1,
\ldots, i_\ell\}\subset \{1, \ldots, s\}$ and
$\sigma = (\sigma_1, \ldots, \sigma_s)\in \{-1, 1\}^s$, the algebraic
sets $V^{\mathcal{I}, \sigma}_{\bm{a}, e}\subset \CC^t$ defined by
  \[
    g_{i_1}+ \sigma_{i_1} a_{i_1} e = \cdots =
g_{i_\ell}+\sigma_{i_\ell} a_{i_\ell}e = 0
  \]
are, either empty, or $(t-\ell)$-equidimensional and smooth, and the
ideal generated by their defining equations is radical.

Note that by the transfer principle, one can choose instead of a
scalar $e$ an infinitesimal $\varepsilon$ so that the algebraic sets
$V^{\mathcal{I}, \sigma}_{\bm{a}, \varepsilon}$ and their defining set
of equations satisfy the above properties. When, in the above
equations, one leaves $\varepsilon$ as a variable, one obtains
equations defining an algebraic set in $\CC^{t+1}$.  We denote by
$\mathfrak{V}^{\mathcal{I}, \sigma}_{\bm{a}, \varepsilon}$ the union
of the $(t+1-\ell)$-equidimensional components of this algebraic set.

Further we also assume that the $a_i$'s are chosen positive.

Denote by $\mathcal{S}^{(\varepsilon)}$ the extension of the
semi-algebraic set $\mathcal{S}$ to $\RR\langle \varepsilon \rangle^t$
; similarly, the extension of any connected component $C$ of
$\mathcal{S}$ to $\RR\langle \varepsilon \rangle^t$ is denoted by
$C^{(\varepsilon)}$.

Now, remark that any connected component $C^{(\varepsilon)}$ of
$\mathcal{S}^{(\varepsilon)}$ contains a connected component of the
semi-algebraic set $\mathcal{S}_{\bm{a}}^{(\varepsilon)}$ defined by:
  \[
    \left ( -a_1\varepsilon \geq g_1 \vee g_1\geq a_1\varepsilon \right )
    \wedge \cdots \wedge
    \left ( -a_s\varepsilon \geq g_s \vee g_s\geq a_s\varepsilon \right )    
  \]
  Hence, we are led to compute sample points per connected component of
  $\mathcal{S}_{\bm{a}}^{(\varepsilon)}$. These will be encoded with
  zero-dimensional parametrizations with coefficients in $\QQ[\varepsilon]$.

By~\cite[Proposition~13.1]{BPR}, in order to compute sample points per
connected component in $\mathcal{S}_{\bm{a}}^{(\varepsilon)}$, it
suffices to compute sample points in the real algebraic sets
$V^{\mathcal{I}, \sigma}_{\bm{a}, \varepsilon}\cap \RR^t$. To do that,
since the algebraic sets $V^{\mathcal{I}, \sigma}_{\bm{a},
\varepsilon}$ satisfy the above regularity properties, we can use the
algorithm and geometric results of~\cite{SaSc03}. To state these
results, one needs to introduce some notation.

Let $\mathfrak{Q}$ be a real field, $\mathfrak{R}$ be a real closure
of $\mathfrak{Q}$ and $\mathfrak{C}$ be an algebraic closure of
$\mathfrak{R}$. For an algebraic set $V\subset \mathfrak{C}^t$ defined
by $h_1=\cdots=h_\ell=0$ ($h_i\in \mathfrak{Q}[\bm{y}]$ with $\bm{y} =
(y_1, \ldots, y_t)$) and $M\in \mathrm{GL}_t(\mathfrak{R})$, we denote
by $V^{M}$ the set $\{M^{-1}\cdot \x\mid \x\in V\}$ and, for $1\leq i
\leq \ell$, by ${h_i}^M$ the polynomial $h_i(M\cdot \bm{y})$ and by
$\pi_i$ the canonical projection $(y_1, \ldots, y_t) \mapsto (y_1, \ldots,
y_i)$ ($\pi_0$ will simply denote $(y_1, \ldots, y_t) \mapsto
\{\bullet\}$). By slightly abusing notation, we will also denote by
$\pi_i$ projections from $\mathfrak{V}^{\mathcal{I}, \sigma}_{\bm{a},
\varepsilon}$ to the first $i$ coordinates $(y_1, \ldots, y_i)$.

We will consider the set of critical points of the restriction of
$\pi_i$ to $V$ and will denote this set by $\crit(\pi_i, V)$ for
$1\leq i \leq \ell$. By \cite[Theorem~2]{SaSc03}, for a generic choice
of ${M}\in \textrm{GL}_t(\mathfrak{R})$, the union of $V^{{M}}\cap
\pi_{t-\ell}^{-1}(0)$ with the sets $\textrm{crit}(\pi_i, V^{{M}})\cap
\pi_{i-1}^{-1}(0)$ (for $1\leq i \leq t-\ell$) is finite and meets all
connected components of $V^{{M}}\cap \mathfrak{R}^t$. Because $V$
satisfies the aforementioned regularity assumptions,
$\textrm{crit}(\pi_i, V^{{M}})\cap \pi_{i-1}^{-1}(0)$ is defined as
the projection on the $\bm{y}$-space of the solution set to the
polynomials
  \[
    \bm{h}^M, \quad (\lambda_1, \ldots, \lambda_\ell).jac(\bm{h}^M, i), \quad
    u_1\lambda_1+\cdots+u_\ell\lambda_\ell = 1, \quad y_1 = \cdots = y_{i-1} = 0,
  \]
where $\bm{h} = (h_1, \ldots, h_\ell)$, $\lambda_1, \ldots,
\lambda_\ell$ are new variables (called Lagrange multipliers),
$jac(\bm{h}^M, i)$ is the Jacobian matrix associated to $\bm{h}^M$
truncated by forgetting its first first $i$ columns and the $u_i$'s
are generically chosen (see also \cite[App. B]{SaSc17}).

Assume that $D$ is the maximum degree of the $h_j$'s and let $E$ be
the length of a straight-line program evaluating $\bm{h}$. Observe now
that, setting the $y_j$'s to $0$ (for $1\leq j \leq i-1$), and using
\cite[Theorem 1]{SaSc18} combined with the degree estimates in
\cite[Section 5]{SaSc18}, we obtain that such systems can be solved
using
  \[
    O\left (\left (\binom{t-i}{\ell} D^\ell(D-1)^{t-(i-1)-\ell}\right )^2
      (E+(t+\ell)D+(t+\ell)^2)(t+\ell)
    \right )
  \]
arithmetic operations in $\mathfrak{Q}$ and have at most 
\[
  \binom{t-i}{\ell}  D^\ell (D-1)^{t-(i-1)-\ell}
\] solutions. 

Going back to our initial problem, one then needs to solve polynomial
systems which encode the set $\crit(\pi_i, V^{\mathcal{I},
\sigma}_{\bm{a}, \varepsilon})$ of critical points of the restriction
of $\pi_i$ to $V^{\mathcal{I}, \sigma}_{\bm{a}, \varepsilon}$. Note
that these systems have coefficients in $\QQ[\varepsilon]$. To solve
such systems, we rely on~\cite{Sc03}, which consists in specializing
$\varepsilon$ to a generic value $v\in \QQ$ and compute a
zero-dimensional parametrization of the solution set to the obtained
system (within the above arithmetic complexity over $\QQ$) and next
use Hensel lifting and rational reconstruction to deduce from this
parametrization a zero-dimensional parametrization with coefficients
in $\QQ(\varepsilon)$. By~\cite[Corollary 1]{Sc03} and
multi-homogeneous bounds on the degree of the critical points of
$\pi_i$ to $\mathfrak{V}^{\mathcal{I}, \sigma}_{\bm{a}, \varepsilon}$
as in \cite[Section~5]{SaSc18}, this lifting step has a cost
\[O\ {\widetilde{~}}\left ( ((t+\ell)^4+ (t+\ell+1)E) \left
(\binom{t-i}{\ell} D^\ell(D-1)^{t-(i-1)-\ell}\right )^2 \right ).\]
  Hence, all in all computing one zero-dimensional parametrization for one
  critical locus uses 
  \[
    O\ {\widetilde{~}}\left ( ((t+\ell)^4D+ (t+\ell+1)E)
      \left (\binom{t-i}{\ell} D^\ell(D-1)^{t-(i-1)-\ell}\right )^2 \right )
  \]
arithmetic operations in $\QQ$. Note that, following \cite{Sc03}, the
degrees in $\varepsilon$ of the numerators and denominators of the
coefficients of these parametrizations are bounded by
$\binom{t}{\ell}D^\ell(D-1)^{t-\ell}$.

Summing up for all critical loci and using
  \[
    \sum_{i=0}^{t-\ell} \binom{t-i}{\ell} = \binom{t+1}{\ell+1},
  \]
the computation for a fixed ${V}^{\mathcal{I}, \sigma}_{\bm{a},
  \varepsilon}$ uses
  \[
    O\ {\widetilde{~}}\left ( ((t+\ell)^4D+ (t+\ell+1)E )
      \binom{t+1}{\ell+1}^2\left ( D^\ell(D-1)^{t-\ell}\right )^2 \right )
  \]
arithmetic operations in $\QQ$. Also, the number of points computed this
way is dominated by 
\[\binom{t+1}{\ell + 1}\left( D^\ell (D-1)^{t-\ell} \right).\] 
Note that the above quantity is upper bounded by $(2D)^t$ and bounds the
  degree of the output zero-dimensional parametrizations.

Taking the sum for all possible algebraic sets ${V}^{\mathcal{I},
\sigma}_{\bm{a}, \varepsilon}$ and remarking that
  \begin{itemize}
  \item the sum of number of indices of cardinality $\ell$ for $0 \leq
\ell \leq t$ is bounded by $s^t$;
  \item the number of sets $\sigma$ for a given $\ell$ is bounded by
$2^t$;
  \item the sum $\sum_{\ell=0}^t \binom{t+1}{\ell+1}^2$ equals
$2\binom{2t+1}{t}-1$
  \end{itemize}
one deduces that all these zero-dimensional parametrizations can be
computed within
  \[
    O\ {\widetilde{~}}
    \left (
      s^t 2^t\binom{2t+1}{t} \left ((2t)^4D+ (2t+1)\Gamma \right)\ D^{2t}
    \right )
  \]
 arithmetic operations in $\QQ$ (recall that $\Gamma$ bounds the
length of a straight line program evaluating all the polynomials
defining our semi-algebraic set $\mathcal{S}$) which we simplify to
  \[
    O\ {\widetilde{~}}
    \left (
      \Gamma\ s^t\ 2^{3t}\ D^{2t+1}
    \right ). 
  \]

Similarly, using the above simplifications, the total number of points
encoded by these zero-dimensional parametrizations is bounded above by
$(2sD)^t$.

At this stage, we have just obtained zero-dimensional
parametrizations with coefficients in $\QQ(\varepsilon)$.

The above bound on the number of returned points is done but it remains
to show how to specialize $\varepsilon$ in order to get sample points
per connected components in $\mathcal{S}$. To do that, given a
parametrization $\mathscr{R}_\varepsilon = (w, v_1, \ldots, v_t)
\subset \QQ(\varepsilon)[u]^{t+1}$, we need to find a specialization
value $e$ for $\varepsilon$ to obtain a parametrization
$\mathscr{R}_e$ such that

\begin{itemize}
  \item the number of real roots of the zero set associated to
$\mathscr{R}_e$ is the same as the number of real roots of the zero
set associated to $\mathscr{R}_\varepsilon$;
  \item when $\eta$ ranges over the interval $]0, e]$ the signs of the
$g_i$'s at the zero set associated to $\eta$ does not vary.
  \end{itemize}
To do that, it suffices to choose $e$ such that it is smaller than the
smallest positive root of the resultant associated to $\left (w,
\frac{\partial w}{\partial u}\right )$ and the smallest positive roots
of the resultant associated to $w$ and $g_i\left (\frac{v_1}{\partial
w / \partial u}, \ldots, \frac{v_t}{\partial w / \partial u}\right
)$. The algebraic cost (i.e. the resultant computations) are dominated
by the complexity estimates of the previous step.

Finally, note that $\Gamma$ can be bounded by $s~\binom{D+t}{t}$ when
the $g_i$'s are given in an expanded form in the monomial
basis. Therefore, the arithmetic complexity for computing sample
points of the semi-algebraic set defined by $g_1\ne 0 ,\ldots,g_s\ne 0$
can be bounded by
  \[
    O\ {\widetilde{~}}
    \left (
      \binom{D+t}{t}\ s^{t+1}\ 2^{3t}\ D^{2t+1}
    \right ). 
  \]
\end{proof}

\begin{remark}
Observe that since the coefficients of the rational parametrizations
with coefficients in $\QQ[\varepsilon]$ have bit size depending both
on the maximum bit size $\tau$ of the coefficients of the input
polynomials $g_1, \ldots, g_s$ and the bit size of the generically
chosen $a_i$'s.
    
    When substituting $\varepsilon$ by a small enough rational number
$e$, one obtains zero-dimensional parametrizations with coefficients
in $\QQ$ of bit size depending on the one of $e$ also. Admissible
values for $e$ depend on the magnitude of the real roots of the
univariate resultant we exhibit in the above proof. Because we start
with rational parametrizations of degree bounded by $O(D)^t$, assuming
that the bit size of the $a_i$'s is bounded by $O(D)^t$ (following
reasonings like the one in \cite{EGS20}), one could show using
standard quantitative results that the bit size of $e$ may be $\tau \
D^{O(t)}$ (because $e$ is obtained through the isolation of real roots
of a univariate polynomial of degree $D^{O(t)}$). However, this is a
worst case analysis and most of the time, we observe in practice that
one can choose for $e$ values of reasonable bit size.
\end{remark}

We end this section with a Corollary which is a consequence of
the proof of \cite[Theorem 13.18]{BPR}. Basically, once we have the
parametrizations computed by the algorithm on which
Theorem~\ref{thm:sample-points} relies, one can compute sample
points per connected components of the semi-algebraic set $\mathcal{S}$
within the same arithmetic complexity bounds. The idea is just to evaluate
the $g_i$'s at these rational parametrizations and use bounds on the
minimal distance between two roots of a univariate polynomial such as
\cite[Prop. 10.22]{BPR}. Hence, the proof of the corollary below follows
\emph{mutatis mutandis} the same steps as the one of \cite[Theorem
13.18]{BPR}.   

\begin{corollary}
\label{cor:sampling}
Let $(g_1, \ldots, g_s)$ in $\QQ[y_1, \ldots, y_t]$ with $D \geq
\max_{1\leq i \leq s} \deg(g_i)$ and $\mathcal{S}\subset \RR^t$ be the
semi-algebraic set defined by
\[
g_1\neq 0, \ldots, g_s\neq 0.
\]
There exists a probabilistic algorithm which on input $(g_1, \ldots,
g_s)$ outputs a finite set of points $\mathscr{P}$ in $\QQ^t$ of
cardinality at most $\left( 2sD \right)^t $ points such that
$\mathscr{P}$ meets every connected component of $\mathcal{S}$ using
  \[
    O\ {\widetilde{~}}
    \left (
      \binom{D+t}{t} s^{t+1} 2^{3t}  D^{2t+1}
    \right ). 
  \]
arithmetic operations in $\QQ$. 
\end{corollary}

Note that the main difference, by contrast with
Theorem~\ref{thm:sample-points}, the above Corollary shows how to
obtain output points with coordinates in $\QQ$.


%% file: hermite.tex
\section{Parametric Hermite matrices}
\label{section:ParamHermite}
In this section, we adapt the construction encoding Hermite's
quadratic forms, also known as Hermite matrices to the context of
parametric systems and describe an algorithm for computing those {\em
parametric Hermite matrices}.
\subsection{Definition}
\label{ssection:construction}
Let $\KK$ be a field and $I\subset \KK[\bx]$ be a zero-dimensional
ideal. Recall that the quotient ring $A_{\KK} = \KK[\bx]/I$ is a
$\KK$-vector space of finite dimension \cite[Section 5.3, Theorem
6]{CLO}.
For $p\in \KK[\bx]$, we denote by $\cL_p$ the multiplication map
$\overline{q} \in A_{\KK} \mapsto \overline{p\cdot q},\in A_{\KK}$. 


Note that the map $\cL_p$ is an endomorphism of $A_{\KK}$ as a
$\KK$-vector space. The Hermite quadratic form associated to $I$ is
defined as the bilinear form that sends
$(\overline{p},\overline{q})\in A_{\KK}\times A_{\KK}$ to the trace of
$\cL_{p\cdot q}$ as an endomorphism of $A_{\KK}$.

We refer to \cite[Chap. 4]{BPR} for more details about Hermite
quadratic forms.


Now, let $\ff=(f_1,\ldots,f_m)$ be a polynomial sequence in
$\QQ[\by][\bx]$. We take the rational function field $\QQ(\by)$ as the
base field $\KK$ and denote by $\langle \ff \rangle_{\KK}$ the ideal
generated by $\ff$ in $\KK[\bx]$. We require that the system $\ff$
satisfies Assumption \eqref{assumption:A}.

This leads to the following well-known lemma, which is the foundation
for the construction of our parametric Hermite matrices.
\begin{lemma}
  \label{lemma:zerodim}
  Assume that $\ff$ satisfies Assumption \eqref{assumption:A}. Then the
ideal $\langle \ff\rangle_{\KK}$ is zero-dimensional.
\end{lemma}
\begin{proof}
Assume that there exists a coordinate $x_i$ for $1\leq i \leq n$
such that $\langle \ff \rangle \cap \CC[\by,x_i] = \langle
0\rangle$. We denote respectively by $\pi_{i}$ and $\tilde{\pi}_{i}$
the projections $(\by,\bx) \mapsto (\by,x_i)$ and $(\by,x_i)\mapsto
\by$. By the assumption above, $\overline{\pi_{i}(\cV)}$ is the whole
space $\CC^{t+1}$. Then, we have the identity
\[\CC^{t+1}=\overline{\left( \tilde{\pi_i}^{-1}(\mathcal{O})\cup
\tilde{\pi_i}^{-1}(\CC^t\setminus\mathcal{O}) \right)\cap \pi_i(\cV)
},\]
where $\mathcal{O}$ be the dense Zariski open subset of $\CC^t$
required in Assumption~\eqref{assumption:A}.

Since $\tilde{\pi}_i$ is a map from $\CC^{t+1}$ to $\CC^t$, its fibers
are of dimension at most $1$. Therefore, we have that $\dim
\tilde{\pi_i}^{-1}(\CC^t\setminus\mathcal{O})\leq
1+\dim(\CC^t\setminus \mathcal{O}) \leq t$. As Assumption
\eqref{assumption:A} holds and $\dim \tilde{\pi}_i^{-1}(\CC^t\setminus
\mathcal{O}) \leq t$, we have that $\dim
\overline{\tilde{\pi_i}^{-1}(\mathcal{O}) \cap \pi_i(\cV)}=t$.
This contradicts to the identity above. We conclude that, for $1
\leq i \leq n$, $\langle \ff \rangle \cap \CC[\by,x_i] \ne \langle 0
\rangle$.

On the other hand, by Assumption \eqref{assumption:A}, the
Zariski-closure of $\pi(\cV)$ is the whole parameter space
$\CC^t$. Thus, we have that $\langle \ff \rangle \cap \CC[\by] =
\langle 0 \rangle$. Since $\langle \ff \rangle \cap \CC[\by] =
(\langle \ff \rangle \cap \CC[\by,x_i])\cap \CC[\by]$ for every $1\leq
i \leq n$, there exists a polynomial $p_i\in \langle \ff \rangle \cap
\CC[\by,x_i]$ whose degree with respect to $x_i$ is non-zero. Clearly,
$p_i$ is an element of the ideal $\langle \ff\rangle_{\KK}$. Thus,
there exists $d_i$ such that $x_i^{d_i}$ is a leading term in $\langle
\ff \rangle_{\KK}$. Hence, $\langle \ff \rangle_{\KK}$ is a
zero-dimensional ideal.
\end{proof}

Lemma \ref{lemma:zerodim} allows us to apply the construction of Hermite
matrices described in \cite[Chap. 4]{BPR} to parametric systems as
follows.

Since the ideal $\langle \ff \rangle_{\KK}$ is zero-dimensional by
Lemma~\ref{lemma:zerodim}, its associated quotient ring $A_\KK =
\KK[\bx]/\langle \ff \rangle_{\KK}$ is a finite dimensional
$\KK$-vector space. Let $\delta$ denote the dimension of $A_\KK$ as a
$\KK$-vector space.

We consider a basis $B=\{b_1,\ldots,b_{\delta}\}$ of $A_\KK$, where
the $b_i$'s are taken as monomials in the variables $\bx$. Such a
basis can be derived from Gr\"obner bases as follows. We fix an
admissible monomial ordering $\succ$ over the set of monomials in the
variables $\bx$ and compute a Gr\"obner basis $G$ with respect to the
ordering $\succ$ of the ideal $\langle \ff \rangle_{\KK}$. Then, the
monomials that are not divisible by any leading monomial of elements
of $G$ form a basis of $A_{\KK}$.

Recall that, for an element $p\in \KK[\bx]$, we denote by
$\overline{p}$ the class of $p$ in the quotient ring $A_{\KK}$. A
representative of $\overline{p}$ can be derived by computing the
normal form of $p$ by the Gr\"obner basis $G$, which results in a
linear combination of elements of $B$ with coefficients in $\QQ(\by)$.

Assume now the basis $B$ of $A_{\KK}$ is fixed. For any
$p\in\KK[\bx]$, the multiplication map $\cL_p$ is an endomorphism of
$A_{\KK}$. Therefore, it admits a matrix representation with respect
to $B$, whose entries are elements in $\QQ(\by)$. The trace of $\cL_p$
can be computed as the trace of the matrix representing it. Similarly,
the Hermite's quadratic form of the ideal $\langle \ff\rangle_{\KK}$ can
be represented by a matrix with respect to $B$. This leads to the
following definition.

\begin{definition}
Given a parametric polynomial system $\ff=(f_1,\ldots,f_m)\subset
\QQ[\by][\bx]$ satisfying Assumption \eqref{assumption:A}. We fix a
basis $B=\{b_1,\ldots,b_{\delta}\}$ of the vector space $\KK[\bx]/\langle
\ff\rangle_{\KK}$. The parametric Hermite matrix associated to $\ff$ with
respect to the basis $B$ is defined as the symmetric matrix
$H=(h_{i,j})_{1\leq i,j\leq \delta}$ where $h_{i,j} =\trace(\cL_{b_i\cdot b_j})$.
\end{definition}
It is important to note that the definition of parametric Hermite
matrices depends both on the input system $\ff$ and the choice of the
monomial basis $B$.

\subsection{Gr\"obner bases and parametric Hermite matrices}

In the previous subsection, we have defined parametric Hermite
matrices assuming one knows a Gr\"obner basis $G$ with respect to some
monomial ordering of the ideal $\langle \ff \rangle_{\KK}$ where
$\KK=\QQ(\by)$ and $\langle \ff \rangle_{\KK}$ is the ideal of
$\KK[\bx]$ generated by $\ff$.

Computing such a Gr\"obner basis may be costly as this would require
to perform arithmetic operations over the field $\QQ(\by)$ (or
$\ZZ/p\ZZ(\by)$ where $p$ is a prime when tackling this computational
task through modular computations). In this paragraph, we show that
one can obtain parametric Hermite matrices by considering some
Gr\"obner bases of the ideal $\langle \ff \rangle\subset \QQ[\by,
\bx]$ (hence, enabling the use of efficient implementations of
Gr\"obner bases such as the $F_4/ F_5$ algorithms \cite{F4,F5}).

Since the graded reverse lexicographical ordering (\emph{grevlex} for
short) is known for yielding Gr\"obner bases of relatively small
degree comparing to other orders, we prefer using this ordering to
construct our parametric Hermite matrices.  Further, we will use the
notation $\DRL(\bx)$ for the grevlex ordering among the variables
$\bx$ (with $x_1\succ \cdots \succ x_n$) and $\DRL(\bx) \succ
\DRL(\by)$ (with $y_1\succ \cdots \succ y_t$) for the elimination
ordering. We denote respectively by $\lm_{\bx}(p)$ and $\lc_{\bx}(p)$
the leading monomial and the leading coefficient of $p \in \KK[\bx]$
with respect to the ordering $\DRL(\bx)$.

\begin{lemma}
  \label{lemma:GB}
Let $\cG$ be the reduced Gr\"obner basis of $\langle \ff\rangle$ with
respect to the elimination ordering $\DRL(\bx) \succ \DRL(\by)$. Then
$\cG$ is also a Gr\"obner basis of $\langle \ff \rangle_{\KK}$ with
respect to the ordering $\DRL(\bx)$.
\end{lemma}
\begin{proof}
Since $\cG$ is a Gr\"obner basis of the ideal $\langle \ff\rangle$,
every polynomial $f_i$ of $\ff$ can be written as $f_i=\sum_{g\in \cG}
c_g\cdot g$ where $c_g \in \QQ[\bx,\by]$. Therefore, any element of
$\langle \ff\rangle_{\KK}$ can also be written as a combination of
elements of $\cG$ with coefficients in $\QQ(\by)[\bx]$. In other
words, $\cG$ is a set of generators of $\langle \ff\rangle_{\KK}$.

Let $p$ be a polynomial in $\KK[\bx]$, $p$ is contained in $\langle
\ff \rangle_{\KK}$ if and only if there exists a polynomial $q\in
\QQ[\by]$ such that $q \cdot p \in \langle \ff \rangle$. Thus, the
leading monomial of $p$ as an element of $\KK[\bx]$ with respect to
the grevlex ordering $\DRL(\bx)$ is contained in the ideal $\langle
\lm_{\bx}(g)\; |\; g \in \cG\rangle$. Therefore, $\cG$ is a Gr\"obner
basis of $\langle \ff \rangle_{\KK}$.
\end{proof}

Hereafter, we denote by $\cG$ the reduced Gr\"obner basis of $\langle
\ff\rangle$ with respect to the elimination ordering $\DRL(\bx) \succ
\DRL(\by)$. Let $\mathcal{B}$ be the set of all monomials in $\bx$
that are not reducible by $\cG$, which is finite by
Lemmas~\ref{lemma:zerodim} and \ref{lemma:GB}. The set $\mathcal{B}$
actually forms a basis of the $\KK$-vector space $\KK[\bx]/\langle \ff
\rangle_{\KK}$. Then, we denote by $\mathcal{H}$ the parametric
Hermite matrix associated to $\ff$ with respect to this basis
$\mathcal{B}$.

We consider the following assumption on the input system $\ff$.
\begin{assumption}
  For $g\in \cG$, the leading coefficient ${\rm
    lc}_{\bx}(g)$ does not depend on the parameters $\by$.
  \label{assumption:C}
\end{assumption}
As the computations in the quotient ring $A_{\KK}$ are done through
normal form reductions by $\cG$, the lemma below is straight-forward.
\begin{lemma}
  \label{lemma:monic}
Under Assumption~\eqref{assumption:C}, the entries of the parametric
Hermite matrix $\cH$ are elements of $\QQ[\by]$.
\end{lemma}
\begin{proof}
Since Assumption \eqref{assumption:C} holds, the leading coefficients
$\lc_{\bx}(g)$ do not depend on parameters $\by$ for $g\in \cG$. The
normal form reduction in $A_{\KK}$ of any polynomial in
$\QQ[\by][\bx]$ returns a polynomial in $\QQ[\by][\bx]$. Thus, each
normal form can be written as a linear combination of $\mathcal{B}$
whose coefficients lie in $\QQ[\by]$. Hence, the multiplication map
$\cL_{b_i\cdot b_j}$ for $1\leq i,j \leq \delta$ can be represented by
polynomial matrices in $\QQ[\by]$ with respect to the basis
$\mathcal{B}$. As an immediate consequence, the entries of $\cH$, as
being the traces of those multiplication maps, are polynomials in
$\QQ[\by]$.
\end{proof}
The next proposition states that Assumption \eqref{assumption:C} is
satisfied by a generic system $\ff$. It implies that the entries of
the parametric Hermite matrix of a generic system with respect to the
basis $\mathcal{B}$ derived from $\cG$ completely lie in
$\QQ[\by]$. We postpone the proof of Proposition \ref{lemma:genericC}
to Subsection \ref{ssection:deg-generic} where we prove a more general
result (see Proposition~\ref{lemma:E-generic}).

\begin{proposition}
  \label{lemma:genericC}
Let $\CC[\bx,\by]_d$ be the set of polynomials in $\CC[\bx,\by]$
having total degree bounded by $d$. There exists a non-empty Zariski
open subset $\mathscr{F}_C$ of $\CC[\bx,\by]_d^n$ such that
Assumption \eqref{assumption:C} is satisfied by any $\ff\in
\mathscr{F}_C \cap \QQ[\bx,\by]^n$.
\end{proposition}

\subsection{Specialization property of parametric Hermite matrices}
\label{ssection:specialization}
Recall that $\cG$ is the reduced Gr\"obner basis of $\langle
\ff\rangle$ with respect to the ordering $\DRL(\bx) \succ \DRL(\by)$
and $\mathcal{B}$ is the basis of $\KK[\bx]/\langle \ff \rangle_{\KK}$
derived from $\cG$ as discussed in the previous subsection. Then,
$\mathcal{H}$ is the parametric Hermite matrix associated to
$\ff$ with respect to the basis $\mathcal{B}$.

Let $\bma\in \CC^t$ and $\phi_{\bma}:\CC(\by)[\bx] \to \CC[\bx]$,
$p(\by,\bx)\mapsto p(\bma,\bx)$ be the specialization map that
evaluates the parameters $\by$ at $\bma$. Then $\ff(\bma,\cdot) =
(\phi_{\bma}(f_1),\ldots,\phi_{\bma}(f_m))$. We denote by
$\cH({\bma})$ the specialization $(\phi_{\bma}(h_{i,j}))_{1\leq i,j
\leq \delta}$ of $\cH$ at $\bma$.

Recall that, for a polynomial $p \in \CC(\by)[\bx]$, the leading
coefficient of $p$ considered as a polynomial in the variables $\bx$
with respect to the ordering $\DRL(\bx)$ is denoted by
$\lc_{\bx}(p)$. In this subsection, for $p \in \CC[\bx]$, we use
$\lm(p)$ to denote the leading monomial of $p$ with respect to the
ordering $\DRL(\bx)$.

Let $\cW_{\infty}\subset \CC^t$ denote the algebraic
set $\cup_{g\in \cG} \VV({\rm lc}_{\bx}(g))$. In Proposition
\ref{prop:specialization}, we prove that, outside $\cW_{\infty}$, the
specialization $\cH(\bma)$ coincides with the classic Hermite matrix
of the zero-dimensional ideal $\ff(\bma,\cdot) \subset \QQ[\bx]$. This
is the main result of this subsection.

Since the operations over the $\KK$-vector space $A_{\KK}$ rely on
normal form reductions by the Gr\"obner basis $\cG$ of $\langle \ff
\rangle_{\KK}$, the specialization property of $\cH$ depends on the
specialization property of $\cG$. Lemma \ref{lemma:specGB} below,
which is a direct consequence of \cite[Theorem 3.1]{Kal97}, provides
the specialization property of $\cG$. We give here a more elementary
proof for this lemma than the one in \cite{Kal97}.

\begin{lemma}
  \label{lemma:specGB}
Let $\bma\in \CC^t \setminus \cW_{\infty}$. Then the specialization
$\cG(\bma,\cdot)\coloneqq \{\phi_{\bma}(g) \; | \; g\in \cG\}$ is a
Gr\"obner basis of the ideal $\langle \ff(\bma,\cdot)\rangle\subset
\CC[\bx]$ generated by $\ff(\bma,\cdot)$ with respect to the ordering
$\DRL(\bx)$.
\end{lemma}
\begin{proof}
Since $\bma \in \CC^t \setminus \cW_{\infty}$, the leading coefficient
$\lc_{\bx}(g)$ does not vanish at $\bma$ for every $g\in \cG$. Thus,
$\lm_{\bx}(g) = \lm(\phi_{\bma}(g))$.

We denote by $\mathcal{M}$ the set of all monomials in the variables
$\bx$ and
\[\mathcal{M}_{\cG} \coloneqq \{m\in \mathcal{M}\;|\; \exists g\in
\cG\; : \; \lm_{\bx}(g)\text{ divides }m\}=\{m\in \mathcal{M}\;|\;
\exists g\in \cG\; : \; \lm(\phi_{\bma}(g)) \text{ divides }m\}.\]

For any $p \in \langle \ff \rangle \subset \QQ[\bx,\by]$, we prove that
$\lm(\phi_{\bma}(f)) \in \mathcal{M}_G$. If $p$ is identically zero,
there is nothing to prove. So, we assume that $p \ne 0$, $p$ is then
expanded in the form below:
\[p = \sum_{m\in \mathcal{M}_{G}} c_m \cdot m + \sum_{m\in
\mathcal{M}\setminus\mathcal{M}_{G}} c_m \cdot m,\] where the $c_m$'s
are elements of $\QQ[\by]$. Since $p$ is not identically zero, there
exists $m\in \mathcal{M}_{\cG}$ such that $c_m\neq 0$.

Since $\cG$ is a Gr\"obner basis of $\langle \ff \rangle_{\KK}$, any
monomial in $\mathcal{M}_{\cG}$ can be reduced by $\cG$ to a unique
normal form in $\KK[\bx]$. These divisions involve denominators, which
are products of some powers of the leading coefficients of $\cG$ with
respect to the variables $\bx$. We write
\[\NF_{\cG}(p) = \sum_{m\in \mathcal{M}_{\cG}} c_m\cdot
\NF_{\cG}(m) + \sum_{m\in \mathcal{M}\setminus \mathcal{M}_{\cG}}
c_m \cdot m .\]
As $p \in \langle \ff \rangle_{\KK}$, we have that
$\NF_{\cG}(p) = 0$, which implies
\[\sum_{m\in \mathcal{M}\setminus \mathcal{M}_{\cG}} c_m \cdot m =
-\sum_{m\in \mathcal{M}_{\cG}} c_m\cdot \NF_{\cG}(m).\]
Therefore, we have the identity
\[p=\sum_{m \in \mathcal{M}_{\cG}} c_m \cdot (m-\NF_{\cG}(m))\]
Since $\bma$ does not cancel any denominator appearing in
$\NF_{\cG}(m)$, we can specialize the identity above without any
problem:
\[\phi_{\bma}(p)=\sum_{m \in \mathcal{M}_{\cG}} \phi_{\bma}(c_m) \cdot
(m-\phi_{\bma}(\NF_{\cG}(m))).\]

If at least one of the $\phi_{\bma}(c_m)$ does not vanish, then the
leading monomial of $\phi_{\bma}(f)$ is in
$\mathcal{M}_{\cG}$. Otherwise, if all the $\phi_{\bma}(c_m)$ are
canceled, then $\phi_{\bma}(p)$ is identically zero, and there is not
any new leading monomial appearing either. So, the leading monomial of
any $p \in \langle \ff_{\bma}\rangle$ is contained in
$\mathcal{M}_{\cG}$, which means $\cG(\bma,\cdot)$ is a Gr\"obner
basis of $\langle \ff(\bma,\cdot)\rangle$ with respect to $\DRL(\bx)$.
\end{proof}

\begin{proposition}
  \label{prop:specialization}
For any $\bma \in \CC^t \setminus \cW_{\infty}$, the specialization
$\cH(\bma)$ coincides with the classic Hermite matrix of the
zero-dimensional ideal $\langle \ff(\bma,\cdot) \rangle \subset
\CC[\bx]$.
\end{proposition}
\begin{proof}
As a consequence of Lemma \ref{lemma:specGB}, each computation in
$A_{\KK}$ derives a corresponding one in $\CC[\bx]/\langle
\ff(\bma,\cdot)\rangle$ by evaluating $\by$ at $\bma$ in every normal
form reduction by $\cG$. This evaluation is allowed since $\bma$ does
not cancel any denominator appearing during the
computation. Therefore, we deduce immediately the specialization
property of the Hermite matrix.
\end{proof}

Using Proposition \ref{prop:specialization} and \cite[Theorem
4.102]{BPR}, we obtain immediately the following corollary that allows
us to use parametric Hermite matrices to count the root of a
specialization of a parametric system.
\begin{corollary}
  \label{corollary:ranksig}
Let $\bma \in \CC^t\setminus \cW_{\infty}$, then the rank of $H(\bma)$
is the number of distinct complex roots of $\ff(\bma,\cdot)$. When $\bma
\in \RR^t \setminus \cW_{\infty}$, the signature of $H(\bma)$ is the
number of distinct real roots of $\ff(\bma,\cdot)$.
\end{corollary}
\begin{proof}
By Proposition \ref{prop:specialization}, $\cH(\bma)$ is a Hermite
matrix of the zero-dimensional ideal $\langle
\ff(\bma,\cdot)\rangle$. Then, \cite[Theorem 4.102]{BPR} implies that
the rank (resp. the signature) of $\cH(\bma)$ equals to the number of
distinct complex (resp. real) solutions of $\ff(\bma,\cdot)$.
\end{proof}

We finish this subsection by giving some explanation for what happens
above $\cW_{\infty}$, where our parametric Hermite matrix $\cH$ does not
have good specialization property.
\begin{lemma}
  \label{lemma:W_inf}
Let $\cW_{\infty}$ defined as above. Then $\cW_{\infty}$
contains all the following sets:
\begin{itemize}
\item The non-proper points of the restriction of $\pi$ to $\cV$ (see
Section~\ref{section:preliminary} for this definition).
\item The set of points $\bma \in \CC^t$ such that the fiber
  $\pi^{-1}(\bma) \cap \cV$ is infinite.
\item The image by $\pi$ of the irreducible components of $\cV$ whose
dimensions are smaller than $t$.
\end{itemize}
\end{lemma}
\begin{proof}
The claim for the set of non-properness of the restriction of $\pi$ to
$\cV$ is already proven in \cite[Theorem 2]{LaRo07}. We focus on the
two remaining sets.

Using the Hermite matrix, we know that for $\bma \in \CC^t\setminus
\cW_{\infty}$, the system $\ff(\bma,\cdot)$ admits a non-empty finite
set of complex solutions. On the other hand, for any $\bma \in
\CC^t$ such that $\pi^{-1}(\bma)\cap \cV$ is infinite,
$\ff(\bma,\cdot)$ has infinitely many complex solutions. Therefore,
the set of such points $\bma$ is contained in $\cW_{\infty}$.

Let $\cV_{> t}$ be the union of irreducible components of $\cV$ of
dimension greater than $t$. By the fiber dimension theorem
\cite[Theorem 1.25]{Shafa}, the fibers of the restriction of $\pi$ to
$\cV_{>t}$ must have dimension at least one. Similarly, the components
of dimension $t$ whose images by $\pi$ are contained in a Zariski
closed subset of $\CC^t$ also yield infinite fibers. Therefore, as
proven above, all of these components are contained in
$\pi^{-1}(\cW_{\infty})$.

We now consider the irreducible components of dimension smaller than
$t$. Let $\cV_{\ge t}$ and $\cV_{<t}$ be respectively the union of
irreducible components of $\cV$ of dimension at least $t$ and at most
$t-1$. We have that $\cV = \cV_{\ge t} \cup \cV_{< t}$. Let $I \subset
\QQ[\bx,\by]$ denote the ideal generated by $\ff$. Using the primary
decomposition of $I$ (see e.g. \cite[Sec. 4.8]{CLO}), we have that
$I$ is the intersection of two ideals $I_{\ge t}$ and $I_{< t}$ such
that $V(I_{\ge t}) = \cV_{\ge t}$ and $V(I_{< t}) = \cV_{< t}$. We write
\[I = I_{\ge t} \cap I_{< t}.\]
We denote by $R$ the polynomial ring $\QQ(\by)[\bx]$. Then, the above
identity is transferred into $R$:
\[I\cdot R = (I_{\ge t}\cdot R) \cap (I_{<t} \cdot R).\]
Since
$\dim(\overline{\pi(\cV_{<t})}) \leq t-1$, then there exists a
non-zero polynomial $p \in I_{<t}\cap \QQ[\by]$. As $p$ is a unit
in $\QQ(\by)$, the ideal $I_{<t}\cdot R$ is exactly $R$. So,
\[I\cdot R = I_{\ge t} \cdot R.\]
Note that, by Lemma \ref{lemma:GB}, $\cG$ is a Gr\"obner basis of
$I\cdot R$, then it is also a Gr\"obner basis of $I_{\ge t} \cdot
R$. Therefore, the Hermite matrices associated to $I$ and $I_{\ge
t}$ (with respect to the basis derived from $\cG$) coincide. So, for
$\bma \not \in \cW_{\infty}$, the ranks of those matrices are equal
and so are the numbers of complex points in $\pi^{-1}(\bma)\cap \cV$
and $\pi^{-1}(\bma) \cap \cV_{\ge t}$. As $\pi^{-1}(\bma) \cap
\cV_{\ge t} \subset \pi^{-1}(\bma)\cap \cV$, we have that
$\pi^{-1}(\bma)\cap \cV = \pi^{-1}(\bma) \cap \cV_{\ge t}$. This leads
to
\[\pi^{-1}(\CC^t\setminus \cW_{\infty})\cap \cV_{\ge t} =
\pi^{-1}(\CC^t\setminus \cW_{\infty})\cap \cV.\]
Then, $\pi^{-1}(\CC^t\setminus \cW_{\infty}) \cap \cV_{< t} = \emptyset$
or equivalently, $\cV_{<t} \subset \pi^{-1}(\cW_{\infty})$, which
concludes the proof.
\end{proof}
\subsection{Computing parametric Hermite matrices}
\label{ssection:computation}
Given $\ff = (f_1,\ldots,f_m) \in \QQ[\by][\bx]$ satisfying Assumption
\eqref{assumption:A}. We keep denoting $\KK = \QQ(\by)$. Let $\cG$ be
the reduced Gr\"obner basis of $\langle \ff\rangle$ with respect to
the ordering $\DRL(\bx) \succ \DRL(\by)$ and $\mathcal{B}$ be the set
of all monomials in the variables $\bx$ which are not reducible by
$\cG$. The set $\mathcal{B}$ then forms a basis of the $\KK$-vector
space $\KK[\bx] / \langle \ff\rangle_{\KK}$.

In this subsection, we focus on the computation of the parametric
Hermite matrix associated to $\ff$ with respect to the basis
$\mathcal{B}$.

Note that one can design an algorithm using only the definition of
parametric Hermite matrices given in
Subsection~\ref{ssection:construction}. More precisely, for each
$b_i\cdot b_j \in \mathcal{B}$ ($1\leq i, j \leq \delta$), one
computes the matrix representing $\cL_{b_i\cdot b_j}$ in the basis
$\mathcal{B}$ by computing the normal form of every $b_i\cdot b_j\cdot
b_k$ for $1 \leq k \leq \delta$. Therefore, in total, this direct
algorithm requires $O(\delta^3)$ normal form reductions of polynomials
in $\KK[\bx]$.

In Algorithm \ref{algo:DRL-Matrix} below, we present another algorithm
for computing $\cH$. We call to the following subroutines successively:

\begin{itemize}
\item[$\bullet$] {\sf GrobnerBasis} that takes as input the system
$\ff$ and computes the reduced Gr\"obner basis $\cG$ of $\langle \ff
\rangle$ with respect to the ordering $\DRL(\bx) \succ \DRL(\by)$ and
the basis $\mathcal{B} =\{b_1,\ldots,b_{\delta}\}\subset \QQ[\bx]$
derived from $\cG$.

Such an algorithm can be obtained using any general algorithm for
computing Gr\"obner basis, which we refer to F4/F5 algorithms
\cite{F4,F5}.
\item[$\bullet$] {\sf ReduceGB} that takes as input the Gr\"obner
basis $\cG$ and outputs a subset $\cG'$ of $\cG$ which is still a
Gr\"obner basis of $\langle \ff \rangle_{\KK}$ with respect to the
ordering $\DRL(\bx)$.
  
This subroutine aims to remove the elements in $\cG$ that we do not
need. Even though $\cG$ is reduced as a Gr\"obner basis of $\langle
\ff\rangle$ with respect to $\DRL(\bx) \succ \DRL(\by)$, it is not
necessarily the reduced Gr\"obner basis of $\langle \ff\rangle_{\KK}$
with respect to $\DRL(\bx)$. Using \cite[Lemma 3, Sec. 2.7]{CLO}, we
can design {\sf ReduceGB} to remove all the elements of $\cG$ which
have duplicate leading monomials (in $\bx$). We obtain as output a
subset $\cG'$ of $\cG$ which is also a Gr\"obner basis $\cG'$ for
$\langle \ff\rangle_{\KK}$ with respect to $\DRL(\bx)$. Note that this
tweak reduces not only the cardinal of the Gr\"obner basis in use but
also the size of the set $\cW_{\infty}$ introduced in Subsection
\ref{ssection:specialization} (as we have less leading coefficients).

\item[$\bullet$] {\sf XMatrices} that takes as input $(\cG',
\mathcal{B})$ and computes the matrix representation of the
multiplication maps $\cL_{x_i}$ ($1\leq i \leq n$) with respect to
$\mathcal{B}$.
  
This computation is done directly by reducing every $x_i\cdot b_j$
($1\leq i \leq n$, $1\leq j \leq \delta$) to its normal form in
$\KK[\bx]/\langle \ff \rangle_{\KK}$ using $\cG'$.

\item[$\bullet$] {\sf BMatrices} that takes as input the matrices
representing $(\cL_{x_1},\ldots,\cL_{x_n})$ and $\mathcal{B}$ and
computes the matrices representing the $\cL_{b_i}$'s ($1\leq i\leq
\delta$) in the basis $\mathcal{B}$.

We design {\sf BMatrices} in a way that it
constructs the matrices of $\cL_{b_i}$'s inductively in the degree of
the $b_i$'s as follows.

At the beginning, we have the multiplication matrices of $1$ and
the $x_i$'s; those are the matrices of the elements of degree zero and
one. Note that, for any element $b$ of $\mathcal{B}$. At the step of
computing the matrix of an element $b\in \mathcal{B}$, we remark that
there exist a variable $x_i$ and a monomial $b'\in \mathcal{B}$ such
that $b=x_i\cdot b'$ and the matrix of $b'$ is already computed (as
$\deg(b') <\deg(b)$. Therefore, we simply multiply the matrices of
$\cL_{x_i}$ and $\cL_{b'}$ to obtain the matrix of $\cL_{b}$.

\item[$\bullet$] {\sf TraceComputing} that takes as input the
multiplication matrices $\cL_{b_1},\ldots,\cL_{b_{\delta}}$ and
computes the matrix $(\trace(\cL_{b_i\cdot b_j}))_{1\leq i \leq j \leq
\delta}$. This matrix is in fact the parametric Hermite matrix $\cH$
associated to $\ff$ with respect to the basis $\mathcal{B}$. To design
this subroutine, we use the following remark given in \cite{Ro99}.

Let $p,q \in \KK[\bx]$. The normal form $\overline{p}$ of $p$ by $\cG$
can be written as $\overline{p} = \sum_{i=1}^{\delta} c_i\cdot b_i$
where the $c_i$'s lie in $\KK$. Then, we have the identity
\[{\rm trace}(\cL_{p\cdot q})= \sum_{i=1}^{\delta} c_i\cdot
\trace(\cL_{q \cdot b_i}),\]
Hence, by choosing $p = b_i\cdot b_j$ and $q =1$, we can compute
$h_{i,j}$ using the normal form $\overline{b_i\cdot b_j}$ and
$\trace(\cL_{b_1}),\ldots,\trace(\cL_{b_{\delta}})$.

Note that $\trace(\cL_{b_i})$ is easily computed from the matrix of
the map $\cL_{b_i}$. On the other hand, the normal form
$\overline{b_i\cdot b_j}$ can be read off from the $j$-th row of the
matrix representing $\cL_{b_i}$, which is already computed at this
point.

It is also important to notice that there are many duplicated entries
in $\cH$. Thus, we should avoid all the unnecessary
re-computation. This is done easily be keeping a list for tracking
distinct entries of $\cH$.
\end{itemize}

The pseudo-code of Algorithm \ref{algo:DRL-Matrix} is presented
below. Its correctness follows simply from our definition of
parametric Hermite matrices.

Beside the parametric Hermite matrix
$\cH$, we return a polynomial $\bm{w}_{\infty}$ which is the
square-free part of ${\rm lcm}_{g\in \cG}(\lc_{\bx}(g))$ for further
usage. Note that $V(\bm{w}_{\infty}) = \cW_{\infty}$.

\begin{center}
\begin{algorithm}[H]
\KwData{A parametric polynomial system $\ff=(f_1,\ldots,f_m)$}
\KwResult{A parametric Hermite matrix $\cH$ associated to $\ff$ with
  respect to the basis $\mathcal{B}$}
\SetAlgoNoLine
$\cG, \mathcal{B} \gets \textsf{Gr\"obnerBasis}(\ff, \DRL(\bx) \succ
\DRL(\by))$ \\
$\cG' \gets {\sf ReduceGB}(\cG)$ \\
$\bm{w}_{\infty} \gets {\sf sqfree}({\rm lcm}_{g\in \cG}(\lc_{\bx}(g)))$ \\
$(\cL_{x_1},\ldots,\cL_{x_n})\gets {\sf XMatrices}(\cG',\mathcal{B})$ \\
$(\cL_{b_1},\ldots,\cL_{b_{\delta}}) \gets {\sf
BMatrices}((\cL_{x_1},\ldots,\cL_{x_n}),\mathcal{B})$ \\
$\cH \gets {\sf TraceComputing}(\cL_{b_1},\ldots,\cL_{b_{\delta}})$\\
\Return $[\cH,\bm{w}_{\infty}]$
\caption{\textsf{DRL-Matrix}}
\label{algo:DRL-Matrix}
\end{algorithm}
\end{center}

\paragraph{Removing denominators}
Note that, through the computation in the quotient ring $A_{\KK}$, the
entries of our parametric Hermite matrix possibly contains
denominators that lie in $\QQ[\by]$. As the algorithm that we
introduce in Section \ref{section:algorithm} will require us to
manipulate the parametric Hermite matrix that we compute, these
denominators can be a bottleneck to handle the matrix. Therefore, we
introduce an extra subroutine {\sf RemoveDenominator} that returns a
parametric Hermite matrix $\cH'$ of $\ff$ without denominator.

\begin{itemize}
\item[$\bullet$] {\sf RemoveDenominator} that takes as input the
matrix $\cH$ computed by {\sf DRL-Matrix} and outputs a matrix $\cH'$
which is the parametric Hermite matrix associated to $\ff$ with
respect to a basis $\mathcal{B'}$ that will be made explicit below.

As we can freely choose any basis of form $\{c_i\cdot b_i \; | \;1\leq i
\leq \delta\}$ where the $c_i$'s are elements of $\QQ[\by]$, we should
use a basis that leads to a denominator-free matrix. To do this, we
choose $c_i$ as the denominator of $\trace(\cL_{b_i})$ (which lies in
the first row of the matrix $\cH$ computed by {\sf
  TraceComputing}). Then, for the entry of $\cH$ that corresponds to
$b_i$ and $b_j$, we can multiply it with $c_i\cdot c_j$. The output
matrix $\cH'$ is the parametric Hermite matrix associated to $\ff$ with
respect to the basis $\{c_i\cdot b_i\; | \; 1\leq i \leq \delta\}$.

We observe in many examples that this subroutine returns either a
denominator-free matrix or a matrix with smaller degree
denominators. Thus, it facilitates further computations on the output
matrix.
\end{itemize}

\paragraph{Evaluation \& interpolation scheme for generic systems}
Here we assume that the input system $\ff$ satisfies Assumption
\eqref{assumption:C}. By Lemma \ref{lemma:monic}, the entries of $\cH$
are polynomials in $\QQ[\by]$. Suppose that we know beforehand a value
$\Lambda$ that is larger than the degree of any entry of $\cH$, we can
compute $\cH$ by an evaluation \& interpolation scheme as follows.

We start by choosing randomly a set $\mathcal{E}$ of $\binom{t +
\Lambda}{t}$ distinct points in $\QQ^t$. Then, for each $\bma \in
\mathcal{E}$, we use {\sf DRL-Matrix}
(Algorithm~\ref{algo:DRL-Matrix}) on the input $\ff(\bma,\cdot)$ to
compute the classic Hermite matrix associated to $\ff(\bma,\cdot)$
with respect to the ordering $\DRL(\bx)$. These computations involve
only polynomials in $\QQ[\bx]$ and not in $\QQ(\by)[\bx]$. Finally, we
interpolate the parametric Hermite matrix $\cH$ from its specialized
images $\cH(\bma)$ computed previously.

Since Assumption \eqref{assumption:C} holds, then $\cW_{\infty}$ is
empty. By Proposition \ref{prop:specialization}, the Hermite matrix of
$\ff(\bma,\cdot)$ with respect to $\DRL(\bx)$ is the image $\cH(\bma)$
of $\cH$. Therefore, the above scheme computes correctly the
parametric Hermite matrix $\cH$.

We also remark that, in the computation of the specializations
$\cH(\bma)$, we can replace the subroutine {\sf XMatrices} in {\sf
DRL-Matrix} by a linear-algebra-based algorithm described in
\cite{Fa13}. That algorithm constructs the Macaulay matrix and carries
out matrix reductions to obtain simultaneously the normal forms
that {\sf XMatrices} requires.

In Section~\ref{section:complexity}, we will estimate the
complexity of this evaluation \& interpolation scheme when the input
system $\ff$ satisfies some generic assumptions.

%% file: algorithm.tex
\section{Algorithms for real root classification}
\label{section:algorithm}
We present in this section two algorithms targeting the real root classification
problem through parametric Hermite matrices. The one described in Subsection
\ref{ssection:algo-weak} aims to solve the weak version of Problem
\eqref{problem:rrc}. The second algorithm, given in Subsection
\ref{ssection:algo-strong} outputs the semi-algebraic formulas of the cells
$\cS_i$ that solves Problem \eqref{problem:rrc}. Further, in Section
\ref{section:complexity}, we will see that, for a generic sequence $\ff$, the
semi-algebraic formulas computed by this algorithm consist of polynomials of
degree bounded by $n(d-1)d^n$. Up to our knowledge, this improves all previously
known bounds.

Throughout this section, our input is a parametric polynomial system
$\ff=(f_1,\ldots,f_m) \subset \QQ[\by][\bx]$. We require that $\ff$
satisfies Assumptions \eqref{assumption:A} and that the ideal $\langle
\ff \rangle$ is radical.



Let $\cG$ be the reduced Gr\"obner basis of the ideal $\langle \ff
\rangle \subset \QQ[\bx,\by]$ with respect to the ordering $\DRL(\bx)
\succ \DRL(\by)$. Let $\KK$ denote the rational function field
$\QQ(\by)$. We recall that $\mathcal{B}\subset \QQ[\bx]$ is the basis
of $\KK[\bx]/\langle \ff\rangle_{\KK}$ derived from $\cG$ and $\cH$ is
the parametric Hermite matrix associated to $\ff$ with respect to the
basis $\mathcal{B}$.

\subsection{Algorithm for the weak-version of Problem
  \eqref{problem:rrc}}
\label{ssection:algo-weak}
From Subsection \ref{ssection:specialization}, we know that, outside
the algebraic set $\cW_{\infty} \coloneqq \cup_{g\in \cG}
V(\lc_{\bx}(g))$, the parametric matrix $\cH$ possesses good
specialization property (see Proposition~\ref{prop:specialization}).
We denote by $\bm{w}_{\infty}$ the square-free part of ${\rm
lcm}_{g\in \cG}\lc_{\bx}(g)$. This polynomial $\bm{w}_{\infty}$ is
returned as an output of Algorithm~\ref{algo:DRL-Matrix}. Note that
$V(\bm{w}_{\infty}) = \cW_{\infty}$.

\begin{lemma}
  \label{lemma:det-non-zero}
When Assumption~\eqref{assumption:A} holds and the ideal $\langle \ff
\rangle$ is radical, the determinant of $\cH$ is not identically zero.
\end{lemma}
\begin{proof}
Recall that $\KK$ denotes the rational function field $\QQ(\by)$. We
prove that the ideal $\langle \ff \rangle_{\KK} \subset \KK[\bx]$ is
radical.

Let $p \in \KK[\bx]$ such that there exists $n\in \mathbb{N}$
satisfying $p^n \in \langle \ff \rangle_{\KK}$. Therefore, there
exists a polynomial $q \in \QQ[\by]$ such that $q\cdot p^n \in \langle
\ff \rangle$. Then, $(q\cdot p)^n \in \langle \ff \rangle$. As
$\langle \ff \rangle$ is radical, we have that $q\cdot p \in \langle
\ff \rangle$. Thus, $p \in \langle \ff \rangle_{\KK}$, which concludes
that $\langle \ff \rangle_{\KK}$ is radical.

By Lemma \ref{lemma:zerodim}, $\langle \ff\rangle_{\KK}$ is a radical
zero-dimensional ideal in $\QQ(\by)$. Since $\cH$ is also a Hermite
matrix (in the classic sense) of $\langle \ff \rangle_{\KK}$, $\cH$
is full rank. Therefore, $\det(\cH)$ is not identically zero.
\end{proof}

Let $\bm{w}_{\cH} \coloneqq
\mathfrak{n}/\gcd(\mathfrak{n},\bm{w}_{\infty})$ where $\mathfrak{n}$
is the square-free part of the numerator of $\det(\cH)$. We denote by
$\cW_{\cH}$ the vanishing set of $\bm{w}_{\cH}$. By Lemma
\ref{lemma:det-non-zero}, $\cW_{\cH}$ is a proper Zariski closed
subset of $\CC^t$. Our algorithm relies on the following proposition.

\begin{proposition}
  \label{prop:correctness-weak}
Assume that Assumption \eqref{assumption:A} holds and the ideal
$\langle \ff\rangle$ is radical. Then, for each connected component
$\cS$ of the semi-algebraic set $\RR^t \setminus (\cW_{\infty} \cup
\cW_{\cH})$, the number of real solutions of $\ff(\bma,\cdot)$ is
invariant when $\bma$ varies over $\cS$.
\end{proposition}
\begin{proof}
By Lemma \ref{lemma:W_inf}, $\cW_{\infty}$ contains the
following sets:
\begin{itemize}
\item The non-proper points of the restriction of $\pi$ to $\cV$.
\item The point $\bma \in \CC^t$ such that the fiber
  $\pi^{-1}(\bma)\cap \cV$ is infinite.
\item The image by $\pi$ of the irreducible components of $\cV$ whose
dimensions are smaller than $t$.
\end{itemize}


Now we consider the set $K(\pi,\cV) \coloneqq \sing(\cV) \cup
\crit(\pi,\cV)$. Let $\Delta\coloneqq {\rm jac}(\ff,\bx)$ be the
Jacobian matrix of $\ff$ with respect to the variables $\bx$. The
ideal generated by the $n\times n$-minors of $\Delta$ is denoted by
$I_{\Delta}$. Note that, since $\ff$ is radical, $K(\pi,\cV)$ is the
algebraic set defined by the ideal $\langle \ff \rangle + I_{\Delta}$.

By Proposition \ref{prop:specialization}, for $\bma \in \CC^t\setminus
\cW_{\infty}$, $\langle \ff \rangle$ is a zero-dimensional ideal and the
quotient ring $\CC[\bx]/\langle \ff(\bma,\cdot)\rangle$ has dimension
$\delta$. Moreover, if $\bma \in \CC^t \setminus (\cW_{\infty}\cup
\cW_{\cH})$, the system $\ff(\bma,\cdot)$ has $\delta$ distinct complex
solutions as the rank of $\cH(\bma)$ is $\delta$. Therefore,
every complex root of $\ff(\bma,\cdot)$ is of multiplicity one (we use
the definition of multiplicity given in \cite[Sec. 4.5]{BPR}).

Now we prove that, for such a point $\bma$, the fiber $\pi^{-1}(\bma)$
does not intersect $K(\pi,\cV)$. Assume by contradiction that there
exists a point $(\bma,\chi) \in \CC^{t+n}$ lying in $\pi^{-1}(\bma)
\cap K(\pi,\cV)$. Note that $\chi$ is a solution of $\ff(\bma,\cdot)$,
i.e., $\ff(\bma,\chi) = 0$.

As $(\bma,\chi)\in K(\pi,\cV)$, then it is contained in
$V(I_{\Delta})$. Hence, as the derivation in $\Delta$ does not involve
$\by$, $\chi$ cancels all the $n \times n$-minors of the Jacobian
matrix ${\rm jac}(\ff(\bma,\cdot),\bx)$. \cite[Proposition 4.16]{BPR}
implies that $\chi$ has multiplicity greater than one. This
contradicts to the claim that $\ff(\bma,\cdot)$ admits only complex
solutions of multiplicity one.

Therefore, we conclude that, for $\bma \in
\CC^t\setminus (\cW_{\infty} \cup \cW_{\cH})$, $\pi^{-1}(\bma)$ does not
intersect $K(\pi,\cV)$.

So, using what we prove above and Lemma \ref{lemma:W_inf}, we deduce
that, for $\bma \in \RR^t \setminus (\cW_{\infty}\cup \cW_{\cH})$, then
there exists an open neighborhood $O_{\bma}$ of $\bma$ for the
Euclidean topology such that $\pi^{-1}(O_{\bma})$ does not intersect
$K(\pi,\cV) \cup \pi^{-1}(\cW_{\infty})$.

Therefore, by Thom's isotopy lemma \cite{CoSh92}, the projection
$\pi$ realizes a locally trivial fibration over $\RR^t\setminus
(\cW_{\infty}\cup \cW_{\cH})$. So, for any connected component $\cC$ of
$\RR^t \setminus (\cW_{\infty} \cup \cW_{\cH})$ and any $\bma \in \cC$, we
have that $\pi^{-1}(\cC) \cap \cV\cap \RR^{t+n}$ is homeomorphic to
$\cC \times (\pi^{-1}(\bma)\cap \cV\cap \RR^{t+n})$.

As a consequence, the number of distinct real solutions of
$\ff(\bma,\cdot)$ is invariant when $\bma$ varies over each connected
component of $\RR^t \setminus (\cW_{\infty}\cup \cW_{\cH})$.
\end{proof}

To describe Algorithm \ref{algo:Weak-RRC-Hermite}, we need to
introduce the following subroutines:
\begin{itemize}
  \item[$\bullet$] {\sf CleanFactors} which takes as input a polynomial
$p \in \QQ[\by,\bx]$ and the polynomial $\bm{w}_{\infty}$. It computes
the square-free part of $p$ with all the common factors with
$\bm{w}_{\infty}$ removed.

  \item[$\bullet$] {\sf Signature} which takes as input a symmetric
    matrix with entries in $\QQ$ and evaluates its signature.

  \item[$\bullet$] {\sf SamplePoints} which takes as input a set of
polynomials $g_1,\ldots,g_s \in \QQ[\by]$ and computes a finite subset
$\mathcal{R}$ of $\QQ^t$ that intersects every connected component of
the semi-algebraic set defined by $\wedge_{i=1}^s g_i \ne 0$. An
explicit description of {\sf SamplePoints} is given in the proof of
Theorem~\ref{thm:sample-points} in Section~\ref{section:sample-points}.
\end{itemize}

The pseudo-code of Algorithm \ref{algo:Weak-RRC-Hermite} is below. Its
proof of correctness follows immediately from Proposition
\ref{prop:correctness-weak} and Corollary \ref{corollary:ranksig}.

\begin{center}
\begin{algorithm}[H]
\KwData{A polynomial sequence $\ff \in \QQ[\by][\bx]$ such that
$\langle \ff \rangle$ is radical and Assumptions \eqref{assumption:A}
holds.}
\KwResult{A set of sample points and the corresponding numbers of real
  solutions solving the weak version of Problem \eqref{problem:rrc}}
\SetAlgoNoLine
$[\cH, \bm{w}_{\infty}] \gets \textsf{DRL-Matrix}(\ff)$\\
$\bm{w}_{\cH}\gets {\sf CleanFactors}({\rm numer}(\det(\cH)),\bm{w}_{\infty})$\\
$L \gets {\sf SamplePoints}(\bm{w}_{\cH}\ne 0 \wedge \bm{w}_{\infty}
\ne 0)$\\
\For{$\bma \in L$}{
$r_{\bma} \gets {\sf Signature}(\cH(\bma))$
}
\Return $\{(\bma, r_{\bma})\;|\; \bma \in L\}$
\caption{\textsf{Weak-RRC-Hermite}}
\label{algo:Weak-RRC-Hermite}
\end{algorithm}
\end{center}



\begin{remark}
As we have seen, Algorithm \ref{algo:Weak-RRC-Hermite} obtains a
polynomial which serves similarly as discriminant varieties
\cite{LaRo07} or border polynomials \cite{YangXia05} through
computing the determinant of parametric Hermite matrices. Whereas, the
two latter strategies rely on algebraic elimination based on Gr\"obner
bases to compute the projection of $\crit(\pi,\cV)$ on the
$\by$-space. Since it is well-known that the computation of such a
Gr\"obner basis could be heavy, our algorithm has a chance to be more
practical. In Section \ref{section:experiments}, we provide
experimental results to support this claim.
\end{remark}

\begin{remark}
It is worth noticing that, even though the design of Algorithm
\ref{algo:Weak-RRC-Hermite} employs the grevlex monomial ordering
where $x_1 \succ \cdots \succ x_n$, we can replace it by any grevlex
ordering with another lexicographical order among the $\bx$'s. For
instance, we can use the monomial ordering $\DRL(x_n \succ
\cdots \succ x_1)$. While every theoretical claim still holds for this
ordering, the practical behavior could be different. 
\end{remark}





\subsection{Computing semi-algebraic formulas}
\label{ssection:algo-strong}
By Corollary \ref{corollary:ranksig}, the number of real roots of the
system $\ff(\bma,\cdot)$ for a given point $\bma \in \RR^t\setminus
\cW_{\infty}$ can be obtained by evaluating the signature of the
parametric Hermite matrix $\cH$. We recall that the signature of a
matrix can be deduced from the sign pattern of its leading principal
minors. More precisely, we recall the following criterion, introduced
by \cite{Syl1852} and \cite{Jacobi57} (see \cite{GhysRa16} for a
summary on these works).

\begin{lemma}{\cite[Theorem 2.3.6]{GhysRa16}}
  \label{lemma:Hurwitz}
Let $S$ be a $\delta \times \delta$ symmetric matrix in
$\RR^{\delta \times \delta}$ and, for $1\leq i
\leq \delta$, $S_i$ be the $i$-th leading principal minor of $S$, i.e.,
the determinant of the sub-matrix formed by the first $i$ rows and $i$
columns of $S$. By convention, we denote $S_0=1$.

We assume that $S_i\ne 0$ for $0\leq i \leq \delta$. Let $k$ be the
number of sign variations between $S_i$ and $S_{i+1}$. Then, the
numbers of positive and negative eigenvalues of $S$ are respectively
$\delta-k$ and $k$. Thus, the signature of $S$ is $\delta-2k$.
\end{lemma}

This criterion leads us to the following idea. Assume that none of the
leading principal minors of $\cH$ is identically zero. We consider the
semi-algebraic subset of $\RR^t$ defined by the non-vanishing of those
leading principal minors. Over a connected component $\cS'$ of this
semi-algebraic set, each leading principal minor is not zero and its
sign is invariant. As a consequence, by Lemma \ref{lemma:Hurwitz} and
Corollary \ref{corollary:ranksig}, the number of distinct real roots
of $\ff(\bma,\cdot)$ when $\bma$ varies over $\cS'\setminus \cW_{\infty}$
is invariant.

However, this approach does not apply directly if one of the leading
principle minors of $\cH$ is identically zero. We bypass this obstacle
by picking randomly an invertible matrix $A\in {\rm
GL}_{\delta}(\QQ)$ and working with the matrix $\cH_A \coloneqq
A^T\cdot \cH\cdot A$. The lemma below states that, with a generic
matrix $A$, all of the leading principal minors of $\cH_A$ are not
identically zero.

\begin{lemma}
  \label{lemma:random-matrix}
There exists a Zariski dense subset $\mathcal{A}$ of ${\rm
GL}_{\delta}(\QQ)$ such that for $A \in \mathcal{A}$, all of the
leading principal minors of $\cH_A\coloneqq A^T\cdot \cH \cdot A$ are
not identically zero.
\end{lemma}
\begin{proof}
For $1\leq r\leq \delta$, we denote by $\mathfrak{M}_r$ the set of all
$r\times r$ minors of $\cH$.

Let $\bma \in \QQ^t\setminus \cW_{\infty}\cup \cW_{\cH}$. We have that
$\cH(\bma)$ is a full rank matrix in $\QQ^{\delta\times\delta}$ and,
for $A\in {\rm GL}_{\delta}(\RR)$, $\cH_A(\bma) = A^T\cdot \cH(\bma)
\cdot A$.

We prove that there exists a Zariski dense subset $\mathcal{A}$
of ${\rm GL}_{\delta}(\QQ)$ such that, for $A\in \mathcal{A}$, all of
the leading principal minors of $\cH_A(\bma)$ are not zero. Then, as an
immediate consequence, all the leading principal minors of $\cH_A$ are
not identically zero.

We consider the matrix $A = (a_{i,j})_{1\leq i, j \leq \delta}$ where
$\bm{a}=(a_{i,j})$ are new variables. Then, the $r$-th leading
principal minor $M_r(\bm{a})$ of $A^T \cdot \cH(\bma) \cdot A$ can be
written as
\[M_r(\bm{a}) = \sum_{\mathfrak{m} \in \mathfrak{M}_r}
a_{\mathfrak{m}} \cdot \mathfrak{m}(\bma),\]
where the $a_{\mathfrak{m}}$'s are elements of $\QQ[\bm{a}]$.

As $\cH(\bma)$ is a full rank symmetric matrix by assumption, there
exists a matrix $Q \in {\rm GL}_{\delta}(\RR)$ such that $Q^T\cdot
\cH(\bma) \cdot Q$ is a diagonal matrix with no zero on its diagonal.
Hence, the evaluation of $\bm{a}$ at the entries of $Q$ gives
$M_r(\bm{a})$ a non-zero value. As a consequence, $M_r(\bm{a})$ is not
identically zero.

Let $\mathcal{A}_r$ be the non-empty Zariski open subset of ${\rm
GL}_{\delta}(\QQ)$ defined by $M_r(\bm{a}) \ne 0$. Then, the set of
the matrices $A \in \mathcal{A}_r$ such that the $r\times r$ leading
principal minor of $A^T\cdot \cH(\bma)\cdot A$ is not zero.

Taking $\mathcal{A}$ as the intersection of $\mathcal{A}_r$ for $1\leq
r\leq \delta$, then, for $A \in \mathcal{A}$, none of the leading
principal minors of $A^T\cdot \cH(\bma) \cdot A$ equals
zero. Consequently, each leading principal minor of $A^T \cdot \cH
\cdot A$ is not identically zero.
\end{proof}

Our algorithm (Algorithm \ref{algo:RRC-Hermite}) for solving Problem
\eqref{problem:rrc} through parametric Hermite matrices is described
below. As it depends on the random choice of the matrix $A$,
Algorithm~\ref{algo:RRC-Hermite} is probabilistic. One can easily
modify it to be a Las Vegas algorithm by detecting the
cancellation of the leading principal minors for each choice of $A$.

\begin{center}
\begin{algorithm}[H]
\KwData{A polynomial sequence $\ff \subset \QQ[\by][\bx]$ such that
the ideal $\langle \ff \rangle$ is radical and $\ff$ satisfies
Assumption
\eqref{assumption:A}}
\KwResult{The descriptions of a collection of semi-algebraic sets
$\cS_i$ solving Problem \eqref{problem:rrc}}
\SetAlgoNoLine
$\cH,\bm{w}_{\infty} \gets \textsf{DRL-Matrix}(\ff)$ \label{algo-line:1}\\
Choose randomly a matrix $A$ in $\QQ^{\delta \times \delta}$ \\
$\cH_A \gets A^T\cdot \cH \cdot A$ \\
$(M_1,\ldots,M_{\delta}) \gets {\sf
  LeadingPrincipalMinors}(\cH_A)$ \label{algo-line:2} \\
$L \gets {\sf SamplePoints}\left( \bm{w}_{\infty}\wedge \left(
  \wedge_{i=1}^{\delta} M_i \ne 0 \right) \right)$\\
\For{$\bma \in L$}{
  $r_{\bma} \gets {\sf Signature}(\cH(\bma))$ \\
  }
\Return $\{(\sign(M_1(\bma),\ldots,M_{\delta}(\bma)),\bma,r_{\bma})\;|\;
\bma \in L\}$
\caption{\textsf{RRC-Hermite}}
\label{algo:RRC-Hermite}
\end{algorithm}
\end{center}

\begin{proposition}
  \label{prop:correctness-strong}
Assume that $\ff$ satisfies Assumptions \eqref{assumption:A} and that
the ideal $\langle \ff \rangle$ is radical. Let $A$ be a matrix in
${\rm GL}_{\delta}(\QQ)$ such that all of the leading principal minors
$M_1,\ldots,M_{\delta}$ of $\cH_A\coloneqq A^T\cdot \cH \cdot A$ are
not identically zero. Then, Algorithm \ref{algo:RRC-Hermite} computes
correctly a solution for Problem \eqref{problem:rrc}.
\end{proposition}
\begin{proof}
Note that for $\bma \in \RR^t\setminus \cW_{\infty}$, we have that
$\cH_A(\bma) = A^T\cdot \cH(\bma) \cdot A$. Therefore, the signature of
$\cH(\bma)$ equals to the signature of $\cH_A(\bma)$.

Let $M_1,\ldots,M_{\delta}$ be the leading principal minors of $\cH_A$
and $\cS$ be the algebraic set defined by $\wedge_{i=1}^{\delta} M_i
\ne 0$. Over each connected component $\cS'$ of $\cS$, the sign of each
$M_i$ is invariant and not zero. Therefore, by Lemma
\ref{lemma:Hurwitz}, the signature of $\cH_A(\bma)$, and therefore of
$\cH(\bma)$, is invariant when $\bma$ varies over $\cS' \setminus
\cW_{\infty}$.  As a consequence, by Corollary \ref{corollary:ranksig},
the number of distinct real roots of $\ff(\bma,\cdot)$ is also invariant
when $\bma$ varies over $\cS'\setminus \cW_{\infty}$.  We finish the
proof of correctness of Algorithm \ref{algo:RRC-Hermite}.
\end{proof}

%% file: complexity.tex
\section{Complexity analysis}
\label{section:complexity}
\subsection{Degree bound of parametric Hermite matrices on generic
input}
\label{ssection:deg-generic}
In this subsection, we consider an affine regular sequence $\ff =
(f_1, \ldots, f_n) \subset \QQ[\by][\bx]$ according to the variables
$\bx$, i.e., the homogeneous components of largest degree in $\bx$ of
the $f_i$'s form a homogeneous regular sequence (see
Section~\ref{section:preliminary}). Additionally, we require that $\ff$
satisfies Assumptions~\eqref{assumption:A} and \eqref{assumption:C}.

Let $d$ be the highest value among the total degrees of the
$f_i$'s. Since the homogeneous regular sequences are generic among the
homogeneous polynomial sequences (see, e.g., \cite[Proposition
1.7.4]{Bar-thesis} or \cite{Pardue10}), the same property of
genericity holds for affine regular sequences (thanks to the
definition we use).

As in previous sections, $\cG$ denotes the reduced Gr\"obner basis of
$\langle \ff \rangle$ with respect to the ordering $\DRL(\bx) \succ
\DRL(\by)$. Let $\delta$ be the dimension of the $\KK$-vector space
$\KK[\bx]/\langle \ff\rangle_{\KK}$ where $\KK = \QQ(\by)$. By
B\'ezout's inequality, $\delta \leq d^n$. We derive from $\cG$ a basis
$\mathcal{B} = \{b_1,\ldots,b_{\delta}\}$ of $\KK[\bx]/ \langle \ff
\rangle_{\KK}$ consisting of monomials in the variables
$\bx$. Finally, the parametric Hermite matrix of $\ff$ with respect to
$\mathcal{B}$ is denoted by $\cH = (h_{i,j})_{1\leq i,j \leq \delta}$.

For a polynomial $p\in \QQ[\by,\bx]$, we denote by $\deg(p)$ the total
degree of $p$ in $(\by,\bx)$ and $\deg_{\bx}(p)$ the partial degree of
$p$ in the variables $\bx$.

As Assumption \eqref{assumption:C} holds, by Lemma \ref{lemma:monic},
the entries of the parametric Hermite matrix $\cH$ associated to $\ff$
with respect to the basis $\mathcal{B}$ are elements of $\QQ[\by]$. To
establish a degree bound on the entries of $\cH$, we need to introduce
the following assumption.

\begin{assumption}
  \label{assumption:E}
  For any $g\in \cG$, we have that $\deg(g)=\deg_\bx(g)$.
\end{assumption}
Proposition \ref{lemma:E-generic} below states that Assumption
\eqref{assumption:E} is generic. Its direct consequence is a proof for
Proposition \ref{lemma:genericC}.
\begin{proposition}
  \label{lemma:E-generic}
Let $\CC[\bx,\by]_d$ be the set of polynomials in $\CC[\bx,\by]$
having total degree bounded by $d$. There exists a non-empty Zariski
open subset $\mathscr{F}_D$ of $\CC[\bx,\by]_d^n$ such that Assumption
\eqref{assumption:E} holds for $\ff \in \mathscr{F}_D\cap
\QQ[\bx,\by]^n$.

Consequently, for $\ff \in\mathscr{F}_D \cap \QQ[\bx,\by]^n$, $\ff$
satisfies Assumption \eqref{assumption:C}.
\end{proposition}
\begin{proof}
Let $y_{t+1}$ be a new indeterminate. For any polynomial $p \in
\QQ[\bx,\by]$, we consider the homogenized polynomial $p_h\in
\QQ[\bx,\by,y_{t+1}]$ of $p$ defined as follows:
\[p_h = y_{t+1}^{\deg(p)} p \left(
\frac{x_1}{y_{t+1}},\ldots,\frac{x_n}{y_{t+1}},
\frac{y_1}{y_{t+1}},\ldots,\frac{y_t}{y_{t+1}}\right).\]
Let $\CC[\bx,\by,y_{t+1}]^h_d$ be the set of homogeneous polynomials
in $\CC[\bx,\by,y_{t+1}]$ whose degrees are exactly $d$. By
\cite[Corollary 1.85]{Verron16}, there exists a non-empty Zariski
subset $\mathscr{F}^h_D$ of $\left ( \CC[\bx,\by,y_{t+1}]^h_d \right)
^n$ such that the variables $\bx$ is in Noether position with respect
to $\ff_h$ for every $\ff_h \in \mathscr{F}^h_D$.

For $\ff_h \in \mathscr{F}^h_D$, let $G_h$ be the reduced Gr\"obner
basis of $\ff_h$ with respect to the grevlex ordering $\DRL(\bx \succ
\by \succ y_{t+1})$. By \cite[Proposition 7]{BFS14}, if the variables
$\bx$ is in Noether position with respect to $\ff_h$, then the leading
monomials appearing in $G_h$ depend only on $\bx$.

Let $\ff$ and $G$ be the image of $\ff_h$ and $G_h$ by substituting
$y_{t+1}=1$. We show that $G$ is a Gr\"obner basis of $\ff$ with
respect to the ordering $\DRL(\bx \succ \by)$.

Since $G_h$ generates $\langle \ff_h\rangle$, $G$ is a generating set
of $\langle \ff\rangle$. As the leading monomials of elements in $G_h$
do not depend on $y_{t+1}$, the substitution $y_{t+1}=1$ does not affect
these leading monomials.

For a polynomial $p\in \langle \ff \rangle \subset \QQ[\bx,\by]$, then
$p$ writes $p = \sum_{i=1}^n c_i \cdot f_i$,
where the $c_i$'s lie in
$\QQ[\bx,\by]$. We homogenize the polynomials $c_i\cdot f_i$ on the
right hand side to obtain a homogeneous polynomial $P_h \in \langle
\ff_h\rangle$. Note that $P_h$ is not necessarily the homogenization
$p_h$ of $p$ but only the product of $p_h$ with a power of
$y_{t+1}$. Then, there exists a polynomial $g_h\in G_h$ such that the
leading monomial of $g_h$ divides the leading monomial of $P_h$. Since
the leading monomial of $g_h$ depends only on $\bx$, it also divides
the leading monomial of $p_h$, which is the leading monomial of
$p$. So, the leading monomial of the image of $g_h$ in $G$ divides the
leading monomial of $p$. We conclude that $G$ is a Gr\"obner basis of
$\ff$ with respect to the ordering $\DRL(\bx \succ \by)$ and the set
of leading monomials in $G$ depends only on the variables $\bx$.

Let $\mathscr{F}_D$ be the subset of $\CC[\bx,\by]_d^n$ such that for
every $\ff \in \mathscr{F}_D$, its homogenization $\ff_h$ is contained
in $\mathscr{F}^h_D$. Since the two spaces $\left (
  \CC[\bx,\by,y_{t+1}]_d^h\right )^n$ and
$\CC[\bx,\by]_d^n$ are both exactly $\CC^{\binom{d+n+t}{n+t} \times n}$
(by considering each monomial coefficient as a coordinate),
$\mathscr{F}_D$ is also a non-empty Zariski open subset of
$\CC[\bx,\by]_d^n$.

Assume now that the polynomial sequence $\ff$ belongs to
$\mathscr{F}_D$. We consider the two monomial orderings over
$\QQ[\bx,\by]$ below:
\begin{itemize}
  \item The elimination ordering $\DRL(\bx) \succ
\DRL(\by)$ is abbreviated by $O_1$. The leading monomial of $p\in
\QQ[\bx,\by]$ with respect to $O_1$ is denoted by $\lm_{1}(p)$. The
reduced Gr\"obner basis of $\ff$ with respect to $O_1$ is $\cG$.
  \item The grevlex ordering $\DRL(\bx \succ \by)$ is
abbreviated by $O_2$. The leading monomial of $p\in \QQ[\bx,\by]$ with
respect to $O_2$ is denoted by $\lm_2(p)$. The reduced Gr\"obner basis
of $\ff$ with respect to $O_2$ is denoted by $\cG_2$.
\end{itemize}
As proven above, the set $\{\lm_2(g_2)\; | \; g_2\in
\cG_2\}$ does not depend on $\by$. With this property, we will show,
for any $g_2\in \cG_2$, there exists a polynomial $g \in \cG$ such
that $\lm_{1}(g)$ divides $\lm_{2}(g_2)$.

By definition, $\lm_{2}(g_2)$ is greater than any other monomial of $g_2$
with respect to the ordering $O_2$. Since $\lm_{2}(g_2)$ depends only
on the variables $\bx$, it is then greater than any monomial of $g_2$ with
respect to the ordering $O_1$. Hence, $\lm_{2}(g_2)$ is also
$\lm_{1}(g_2)$. Consequently, since $\cG$ is a Gr\"obner basis of
$\ff$ with respect to $O_1$, there exists a polynomial $g\in \cG$ such
that $\lm_{1}(g)$ divides $\lm_{1}(g_2) = \lm_{2}(g_2)$.

Next, we prove that for every $g \in \cG$, $\lm_1(g)$ is also
$\lm_2(g)$. For this, we rely on the fact that $\cG$ is
reduced. Assume by contradiction that there exists a polynomial $g \in
\cG$ such that $\lm_1(g) \ne \lm_2(g)$. Thus, $\lm_2(g)$ must contain
both $\bx$ and $\by$. Let $t_{\bx}$ be the part in only variables
$\bx$ of $\lm_2(g)$. Note that $\lm_1(g)$ is greater than $t_{\bx}$
with respect to $O_1$. There exists an element $g_2\in \cG_2$ such
that $\lm_2(g_2)$ divides $\lm_2(g)$. Since $\lm_2(g_2)$ depends only
on the variables $\bx$, we have that $\lm_2(g_2)$ divides
$t_{\bx}$. Then, by what we proved above, there exists $g' \in \cG$
such that $\lm_{1}(g)$ divides $\lm_2(g_2)$, so $\lm_1(g)$ divides
$t_{\bx}$. This implies that $\cG$ is not reduced, which contradicts
the definition of $\cG$.

So, $\lm_1(g) = \lm_2(g)$ for every $g \in \cG$ and, consequently,
$\deg(g) = \deg_{\bx}(g)$. We conclude that there exists a non-empty
Zariski open subset $\mathscr{F}_D$ (as above) of $\CC[\bx,\by]_d^n$ such
that Assumption \eqref{assumption:E} holds for every $\ff \in
\mathscr{F}_D\cap \QQ[\bx,\by]^n$.

Additionally, one easily notices that Assumption \eqref{assumption:E}
implies Assumption \eqref{assumption:C}. As a consequence, $\ff$ also
satisfies Assumption \eqref{assumption:C} for any $\ff \in
\mathscr{F}_D\cap \QQ[\bx,\by]^n$.
\end{proof}

Recall that, when Assumption \eqref{assumption:C} holds, by Lemma
\ref{lemma:monic}, the trace of any multiplication map $\cL_p$ is a
polynomial in $\QQ[\by]$ where $p\in \QQ[\by][\bx]$. We now estimate
the degree of $\trace(\cL_p)$. Since the map $p \mapsto \trace(\cL_p)$
is linear, it is sufficient to consider $p$ as a monomial in the
variables $\bx$.

\begin{proposition}
  \label{prop:degree}
Assume that Assumption \eqref{assumption:E} holds. Then, for any
monomial $m$ in the variables $\bx$, the degree in $\by$ of
$\trace(\cL_m)$ is bounded by $\deg(m)$. As a consequence, the total
degree of the entry $h_{i,j}=\trace(\cL_{b_i\cdot b_j})$ of $\cH$ is
at most the sum of the total degrees of $b_i$ and $b_j$, i.e.,
\[\deg(h_{i,j}) \leq \deg(b_i) + \deg(b_j).\]
\end{proposition}
\begin{proof}
Let $m$ be a monomial in $\QQ[\bx]$. The multiplication matrix $\cL_m$
is built as follows. For $1\leq i \leq \delta$, the normal form of
$b_i \cdot m$ as a polynomial in $\QQ(\by)[\bx]$ writes
\[\NF_{\cG}(b_i\cdot m) = \sum_{j=1}^{\delta} c_{i,j}\cdot b_j.\]
Note that this normal form is the remainder of the successive divisions
of $b_i\cdot m$ by polynomials in $\cG$. As Assumption
\eqref{assumption:E} holds, Assumption \eqref{assumption:C} also
holds. Therefore, those divisions do not introduce any
denominator. So, every term appearing during these normal form
reductions are polynomials in $\QQ[\by][\bx]$.

Let $p \in \QQ[\by][\bx]$. For any $g\in \cG$, by Assumption
\eqref{assumption:E}, the total degree in $(\by,\bx)$ of every term of
$g$ is at most the degree of $\lm_{\bx}(g)$. Thus, a division of $p$
by $g$ involves only terms of total degree $\deg(p)$. Thus, during the
polynomial division of $p$ to $\cG$, only terms of degree at most
$\deg(p)$ will appear. Hence the degree of $\NF_{\cG}(p)$ is bounded
by $\deg(p)$.

Note that $\trace(\cL_{m}) = \sum_{i=1}^{\delta} c_{i,i}$. As the
degree of $c_{i,i}\cdot b_i$ is bounded by $\deg(b_i) +\deg(m)$, the
degree of $c_{i,i}$ is at most $\deg(m)$. Then, we obtain that
$\deg(\trace(\cL_m)) \leq \deg(m)$.

Finally, the degree bound of $h_{i,j}$ follows immediately:
\[\deg(h_{i,j}) =\deg(\trace(\cL_{b_i\cdot b_j}))\leq \deg(b_i\cdot
b_j) = \deg(b_i) + \deg(b_j).\]
\end{proof}

\begin{lemma}
  \label{lemma:degree-minor}
Assume that $\ff$ satisfies Assumption \eqref{assumption:E}. Then the
degree of a minor $M$ consisting of the rows $(r_1,\ldots,r_{\ell})$
and the columns $(c_1,\ldots,c_{\ell})$ of $\cH$ is bounded by
\[\sum_{i=1}^{\ell}\left(\deg(b_{r_i})+\deg(b_{c_i})\right).\]
Particularly, the degree of $\det(\cH)$ is bounded by
$2\sum_{i=1}^{\delta} \deg(b_i)$.
\end{lemma}
\begin{proof}
We expand the minors $M$ into terms of the form
$(-1)^{\sign(\sigma)}h_{r_1,\sigma(c_1)}\ldots
h_{r_{\ell},\sigma(c_{\ell})}$, where $\sigma$ is a permutation of
$\{c_1,\ldots,c_{\ell}\}$ and $\sign(\sigma)$ is its signature. We
then bound the degree of each of those terms as follows using
Proposition \ref{prop:degree}:
\[
  \deg \left (\prod_{i=1}^{\ell} h_{r_i,\sigma(c_i)} \right) =
\sum_{i=1}^{\ell} \deg(h_{r_i,\sigma(c_i)})
 \leq \sum_{i=1}^{\ell}\left(\deg(b_{r_i}) +\deg(b_{\sigma(c_i)})\right) = \sum_{i=1}^{\ell}\left(\deg(b_{r_i})+\deg(b_{c_i})\right).
\]
Hence, taking the sum of all those terms, we obtain the inequality:
\[\deg(M_i) \leq
  \sum_{i=1}^{\ell}\left(\deg(b_{r_i})+\deg(b_{c_i})\right).\]
When $M$ is taken as the determinant of $\cH$, then
\[\deg(\det(\cH)) \leq 2\sum_{i=1}^{\delta} \deg(b_i).\]
\end{proof}

Proposition \ref{prop:degree} implies that, when Assumption
\eqref{assumption:E} holds, the degree pattern of $\cH$ depends only
on the degree of the elements of
$\mathcal{B}=\{b_1,\ldots,b_\delta\}$. We rearrange $\mathcal{B}$ in
the increasing order of degree, i.e., $\deg(b_i) \leq \deg(b_j)$ for
$1\leq i < j \leq \delta$. So, $b_1=1$ and $\deg(b_1) = 0$. The degree
bounds of the entries of $\cH$ are expressed by the matrix below
\[
\begin{bmatrix} 
0 & \deg(b_2) & \ldots & \deg(b_{\delta})\\ 
\deg(b_2) & 2\deg(b_2) & \ldots & \deg(b_{\delta}) + \deg(b_2)\\
\vdots & \vdots & \ddots & \vdots \\ 
\deg(b_{\delta}) & \deg(b_{\delta}) + \deg(b_2) &\ldots &
2\deg(b_{\delta})
\end{bmatrix}.
\]

Moreover, using the regularity of $\ff$, we are able to establish
explicit degree bounds for the elements of $\mathcal{B}$ and then, for
the minors of $\cH$.

\begin{lemma}
  \label{lemma:hilbert}
Assume that $\ff$ is an affine regular sequence and let $\mathcal{B}$
be the basis defined as above. Then the highest degree among the
elements of $\mathcal{B}$ is bounded by $n(d-1)$ and
\[2\sum_{i=1}^{\delta} \deg(b_i) \leq n(d-1)d^n.\]
\end{lemma}
\begin{proof}
For $p\in \KK[\bx]$, let $p_h\in
\KK[x_1,\ldots,x_{n+1}]$ be the homogenization of $p$ with respect to
the variable $x_{n+1}$, i.e.,
\[ p_h = x_{n+1}^{\deg_{\bx}(p)} p
  \left(\frac{x_1}{x_{n+1}},\ldots,\frac{x_n}{x_{n+1}}\right).\]
The dehomogenization map ${\rm \alpha}$ is defined as:
\begin{align*}
  \alpha : \; \KK[x_1,\ldots,x_{n+1}] & \to \KK[x_1,\ldots,x_n], \\
  p(x_1,\ldots,x_{n+1}) & \mapsto p(x_1,\ldots,x_n,1).
\end{align*}

Also, the homogeneous component of largest degree of $p$ with respect
to the variables $\bx$ is denoted by ${}^Hp$. Throughout this proof,
we use the following notations:
\begin{itemize}
\item $I = \langle \ff \rangle_{\KK}$ and $\cG$ is
  the reduced Gr\"obner basis of $I$ w.r.t. $\DRL(x_1\succ
  \cdots \succ x_{n})$.
\item $I_h = \langle p_h \; | \; p\in \ff\rangle_{\KK}$ and $\cG_h$ is
  the reduced Gr\"obner basis of $I_h$ w.r.t. $\DRL(x_1\succ
  \cdots \succ x_{n+1})$.
\end{itemize}

The Hilbert series of the homogeneous ideal $I_h$ writes
\[{\rm HS}_{I_h}(z) = \sum_{r = 0}^{\infty} \left (\dim_{\KK} \KK[\bx]_r -
  \dim_{\KK} (I_h \cap \KK[\bx]_r) \right ) \cdot z^r,\]
where $\KK[\bx]_{r} = \{ p \; | \; p\in \KK[\bx]:\ \deg_{\bx}(p) = r
\}$

Since $\ff$ is an affine regular sequence, by definition (see
Section~\ref{section:preliminary}), ${}^H\ff =
({}^Hf_1,\ldots,{}^Hf_n)$ forms a homogeneous regular
sequence. Equivalently, by \cite[Proposition 1.44]{Verron16}, the
homogeneous polynomial sequence $((f_1)_h,\ldots,(f_n)_h,x_{n+1})$ is
regular. Particularly, $((f_1)_h,\ldots, (f_n)_h)$ is a homogeneous
regular sequence and, by \cite[Theorem 1.5]{Moreno03}, we obtain
\[{\rm HS}_{I_h}(z) = \frac{\prod_{i=1}^n \left (1-z^{\deg(f_i)}
\right )}{\left (1-z \right )^{n+1}} = \frac{\prod_{i=1}^n \left
(1+\ldots+z^{\deg(f_i)-1} \right )}{1-z} .\]

On the other hand, as $((f_1)_h,\ldots,(f_n)_h,x_{n+1})$ is a
homogeneous regular sequence, by \cite[Proposition 7]{BFS14}, the
leading terms of $\cG_h$ w.r.t. $\DRL(x_1 \succ \cdots \succ x_{n+1})$
do not depend on the variables $x_{n+1}$. Thus, the dehomogenization
map $\alpha$ does not affect the set of leading terms of
$\cG_h$. Besides, $\alpha(\cG_h)$ is a Gr\"obner basis of $I$ with
respect to $\DRL(\bx)$ (see, e.g., the proof of \cite[Lemma
27]{Spa14}). Hence, the leading terms of $\cG_h$ coincides with the
leading terms of $\cG$.

As a consequence, the set of monomials in $(x_1,\ldots,x_{n+1})$ which
are not contained in the initial ideal of $I_h$ with respect to
$\DRL(x_1 \succ \cdots \succ x_{n+1})$ is exactly
\[\{b \cdot x_{n+1}^j\; | \; b \in \mathcal{B}, j\in \mathbb{N} \}.\]
As a consequence, $\dim_{\KK} \KK[\bx]_r -\dim_{\KK} ( I_h\cap \KK[\bx]_r) =
  \sum_{j=0}^{r} |\mathcal{B} \cap \KK[\bx]_{j}|$.
Let $H(z) = \sum_{r=0}^{\infty} | \mathcal{B} \cap \KK[\bx]_r| \cdot
z^r$. We have that
\[ (1-z)\cdot {\rm HS}_{I_h}(z) = (1-z)\sum_{r=0}^{\infty}
\sum_{j=0}^{r}|\mathcal{B} \cap \KK[\bx]_j| \cdot z^r =
\sum_{r=0}^{\infty} |\mathcal{B} \cap \KK[\bx]_r| \cdot z^r = H(z).\]
Then,
\[H(z) = \prod_{i=1}^n \left ( 1+\ldots+z^{\deg(f_i)-1} \right ).\]
As a direct consequence, $\max_{1\leq i \leq \delta} \deg(b_i)$ is
bounded by $ \sum_{i=1}^n \deg(f_i) - n \leq n(d-1)$.

Let $G_1$ and $G_2$ be two polynomials in $\mathbb{Z}[z]$. We write
$G_1 \leq G_2$ if and only if for any $r \ge 0$, the coefficient of
$z^r$ in $G_2$ is greater than or equal to the one in $G_1$.

Since $\deg(f_i) \leq d$ for every $1 \leq i \leq n$, then
\[H(z) = \prod_{i=1}^n \left ( 1+\ldots+z^{\deg(f_i)-1} \right ) \leq
\prod_{i=1}^n \left ( 1+\ldots +z^{d-1} \right ).\]
As a consequence, $H'(z) = \sum_{r=1}^{\infty} (r\ |\mathcal{B} \cap \KK[\bx]_r |) \cdot
z^{r-1} \leq \left ( \prod_{i=1}^n \left (1+\ldots +z^{d-1} \right )
\right )'$.
Expanding $G'(z)$, we obtain
\begin{align*}
H'(z) & \leq \frac{n \left (\sum_{i=0}^{d-1}z^i \right )^{n-1} \left
(\sum_{i=0}^{d-1}z^i-dz^{d-1} \right )}{1-z} = n \left (\sum_{i=0}^{d-1}z^i \right )^{n-1}\sum_{i=0}^{d-2}z^i
\left (1+\ldots+z^{d-i-2} \right ). 
\end{align*}
By substituting $z=1$ in the above inequality, we obtain
\[H'(1) \leq nd^{n-1}\sum_{i=0}^{d-2}(d-i-1) =
  \frac{n(d-1)d^{n}}{2}.\]
Thus, we have that
$\sum_{i=1}^{\delta}\deg (b_i) = \sum_{r=0}^{\infty} r\ |\mathcal{B}
  \cap \KK[\bx]_r| = H'(1) \leq \frac{n(d-1)d^{n}}{2}$.
\end{proof}

Corollary \ref{corollary:degree-minor} below follows immediately
from Lemmas \ref{lemma:degree-minor} and \ref{lemma:hilbert}.
\begin{corollary}
  \label{corollary:degree-minor}
Assume that $\ff$ is a regular sequence that satisfies
Assumption~\eqref{assumption:E}. Then the degree of any minor of $\cH$
is bounded by $n(d-1)d^n$.
\end{corollary}
\begin{remark}
Note that Assumption \eqref{assumption:E} requires a condition on the
degrees of polynomials in the Gr\"obner basis $\cG$ of $\langle
\ff\rangle$. We remark that it is possible to establish similar bounds
for the degrees of entries of our parametric Hermite matrix and its
minors when the system $\ff$ satisfies a weaker property than
Assumption \eqref{assumption:E} (we still keep the regularity
assumption).

Indeed, we only need to assume that, for any $g\in \cG$, the homogeneous
component of the highest degree in $\bx$ of $g$ does not depend on the
parameters $\by$. Let $d_{\by}$ be an upper bound of the partial
degrees in $\by$ of elements of $\cG$. Under the change of variables
$x_i\mapsto x_i^{d_{\by}}$, $\ff$ is mapped to a new polynomial
sequence that satisfies Assumption \eqref{assumption:E}. Therefore, we
easily deduce the two following bounds, which are similar to the ones
of Proposition \ref{prop:degree} and Corollary
\ref{corollary:degree-minor}.

\begin{itemize}
\item $\deg(h_{i,j}) \leq d_{\by} (\deg(b_i) + \deg(b_j))$;
\item The degree of any minor of $\cH$ is bounded by $d_{\by}~
n(d-1)d^n$.
\end{itemize}
Even though these bounds are not sharp anymore, they still allow us to
compute the parametric Hermite matrices using evaluation \&
interpolation scheme and control the complexity of this computation in
the instances where Assumption \eqref{assumption:E} does not hold.
\end{remark}

\subsection{Complexity analysis of our algorithms}
\label{ssection:complexity}
In this subsection, we analyze the complexity of our algorithms on
generic systems.

Let $\ff = (f_1,\ldots,f_n) \subset \QQ[\bx,\by]$ be a regular
sequence, where $\by = (y_1,\ldots,y_t)$ and $\bx = (x_1,\ldots,x_n)$,
satisfying Assumptions \eqref{assumption:A} and
\eqref{assumption:E}. \revi{To simplify the asymptotic complexity, we
assume that $n$, $t$ and $d$ are greater than or equal to $2$.}

We denote by $\cG$ the reduced Gr\"obner basis of $\ff$ with respect
to the ordering $\DRL(\bx) \succ \DRL(\by)$. The basis $\mathcal{B}$
is taken as all the monomials in $\bx$ that are irreducible by
$\cG$. Then, $\cH$ is the parametric Hermite matrix associated of
$\ff$ with respect to $\mathcal{B}$.
 
We start by estimating the arithmetic complexity for computing the
parametric Hermite matrix $\cH$ and its minors. We denote $\lambda
\coloneqq n(d-1)$ and $\mathfrak{D} \coloneqq n(d-1)d^n$.

\begin{proposition}
  \label{prop:complexity-HM}
Assume that $\ff = (f_1,\ldots,f_n) \subset \QQ[\by][\bx]$ is a
regular sequence that satisfies Assumptions \eqref{assumption:A} and
\eqref{assumption:E}. Let $\delta$ be the dimension of the
$\KK$-vector space $\KK[\bx]/\langle \ff \rangle_{\KK}$ where $\KK =
\QQ(\by)$. Let $\cH$ be the parametric Hermite matrix associated to
$\ff$ constructed using $\DRL(\bx)$ ordering. Then, by Lemma
\ref{lemma:monic}, the entries of the parametric Hermite matrix $\cH$
lie in $\QQ[\by]$.

Using the evaluation \& interpolation scheme, one can compute $\cH$
within
\[O \ {\widetilde{~}} \left(\binom{t+2\lambda}{t}\left (n\
\binom{d+n+t}{n+t} + n^{\omega+1}d^{\omega n +1} + d^{(\omega+1)n}
\right )\right )\]
arithmetic operations in $\QQ$, where, by B\'ezout's bound, $\delta$
is bounded by $d^n$.

Moreover, each minor (including the determinant) of $\cH$ can be
computed using
\[O\ \widetilde{~} \left( \binom{t+\mathfrak{D}}{t} \left ( d^{2n}
\binom{t + 2\lambda}{t} + d^{\omega n} \right) \right )\]
arithmetic operations in $\QQ$.
\end{proposition}
\begin{proof}
By Lemma \ref{lemma:hilbert} and Proposition \ref{prop:degree}, the
highest degree among the entries of $\cH$ is bounded by $2 \lambda =
2n(d-1)$. The evaluation \& interpolation scheme of
Subsection~\ref{ssection:computation} requires computing
$\binom{t+2\lambda}{t}$ specialized Hermite matrices. We first analyze
the complexity for computing each of those specialized Hermite
matrices.

The evaluation of $\ff$ at each point $\bma\in \QQ^t$ costs
$O\left (n\ \binom{d+n+t}{n+t}\right )$ arithmetic operations in $\QQ$.

\revi{As the highest degree in the Gr\"obner basis of
$\ff(\bma,\cdot)$ w.r.t. the $\DRL(\bx)$ ordering is bounded by
$n(d-1)+1$, the computation of this Gr\"obner basis can be done
within $O \left( nd^{\omega n} \right)$ arithmetic operations in $\QQ$
(see \cite[Theorem 5.1]{Fa13}).}

\revi{Next, we compute the matrices representing the
$\cL_{x_i}$'s. Using \cite[Algo. 4]{Fa13}, we obtain an arithmetic
complexity of $O\left(dn^{\omega+1} \delta^{\omega}\right)$
(\cite[Prop. 5]{Fa13}) for computing such $n$ matrices, where $\omega$
is the exponential constant for matrix multiplication. Using $\delta
\leq d^n$, we obtain the bound $O\left(n^{\omega+1}
d^{\omega n+1}\right)$.}

The traces of these matrices are then computed using $n\delta$
additions in $\QQ$. The subroutine {\sf BMatrices} consists of
essentially $\delta$ multiplication of $\delta \times \delta$ matrices
(with entries in $\QQ$). This leads to an arithmetic complexity
$O(\delta^{\omega +1})$, which is then bounded by $O(d^{(\omega
+1)n})$. Next, the computation of each entry $h_{i,j}$ is simply a
vector multiplication of length $\delta$, whose complexity is
$O(\delta)$. Doing so for $\delta^2$ entries, {\sf TraceComputing}
takes in overall $O(\delta^3)$ arithmetic operations in $\QQ$.

Thus, as $\delta \leq d^n$, the complexity of the evaluation step lies
in
\[O\left( \binom{t+2\lambda}{t}\left (n\ \binom{d+n+t}{n+t} +
n^{\omega+1}d^{\omega n + 1} + d^{(\omega+1)n} \right )\right).\]

Finally, we interpolate $\delta^2$ entries which are polynomials in
$\QQ[\by]$ of degree at most $2\lambda$. Using the multivariate
interpolation algorithm of \cite{CaKaYa89}, the complexity of this
step therefore lies in $O\left( \delta^2\
\binom{t+2\lambda}{t} \log^2 \binom{t+2\lambda}{t}
\log\log\binom{t+2\lambda}{t} \right)$.


Summing up the both steps, we conclude that the parametric Hermite
matrix $\cH$ can be obtained within
\[O \ {\widetilde{~}} \left(\binom{t+2\lambda}{t}\left (n\
\binom{d+n+t}{n+t} + n^{\omega+1}d^{\omega n +1} + d^{(\omega+1)n}
\right )\right )\]
arithmetic operations in $\QQ$.

Similarly, the minors of $\cH$ can be computed using the technique of
evaluation \& interpolation. By Corollary
\ref{corollary:degree-minor}, the degree of every minor of $\cH$ is
bounded by $\mathfrak{D}$. We specialize $\cH$ at
$\binom{t+\mathfrak{D}}{t}$ points in $\QQ^t$ and compute the
corresponding minor of each specialized Hermite matrix. This step
takes
\[O\left(\binom{t+\mathfrak{D}}{t} \left ( \delta^2\binom{t + 2\lambda
        }{t}  + \delta^{\omega} \right ) \right)\]
arithmetic operations in $\QQ$. Finally, using the multivariate
interpolation algorithm of \cite{CaKaYa89}, it requires
\[O \left( \binom{t+\mathfrak{D}}{t} \log^2 \binom{t+\mathfrak{D}}{t}
\log\log \binom{t+\mathfrak{D}}{t} \right)\]
arithmetic operations in $\QQ$ to interpolate the final
minor. Therefore, using $\delta \leq d^n$, the whole complexity for
computing each minor of $\cH$ lies within
\[O\ \widetilde{~} \left( \binom{t+\mathfrak{D}}{t} \left ( d^{2n}
\binom{t + 2\lambda}{t} + d^{\omega n} \right) \right ).\]
\end{proof}
\revi{We note that the complexity of computing the matrix
$\mathcal{H}$ in Proposition \ref{prop:complexity-HM}
is also bounded by the complexity of computing its minor. Indeed, we
have that
\begin{align*}
\binom{d+n+t}{n+t} & = \frac{(d+n+t)\ldots(d+n+1)(d+n) \ldots
    (d+1)}{(n+t)!} \\
&\leq \frac{(d+n+t)\ldots(d+n+1)}{t!} \frac{(d+n) \ldots
    (d+1)}{n!}\\
&\leq \frac{(\mathfrak{D}+t) \ldots (\mathfrak{D}+1)}{t!} (2d^n) =
  \binom{\mathfrak{D}+t}{t} (2d^n).
\end{align*}
Asymptotically, $n^{\omega}d^{\omega n +1}$ is bounded by
$O\ \widetilde{~} \left(d^{(\omega+1)n} \right)$. For $t \ge 2$,
$\binom{t+\mathfrak{D}}{t} \ge \mathfrak{D}^2/2 \ge
d^{(\omega-1)n}$. Hence, we obtain
\[\binom{t+2\lambda}{t}\left(n\binom{d+n+t}{n+t} + n^{\omega
+1}d^{\omega n+1}+ d^{(\omega+1)n}\right) \in O\
\widetilde{~}\left(\binom{t+2\lambda}{t}\binom{t+\mathfrak{D}}{t} d^{2
n}\right),\]
which proves our claim above.}

Finally, we state our main result, which is Theorem \ref{thm:main}
below. It estimates the arithmetic complexity of Algorithms
\ref{algo:Weak-RRC-Hermite} and \ref{algo:RRC-Hermite}.

\begin{reptheorem}{thm:main}
Let $\ff \subset \QQ[\bx,\by]$ be a regular sequence such that the
ideal $\langle \ff\rangle$ is radical and $\ff$ satisfies Assumptions
\eqref{assumption:A} and \eqref{assumption:E}. Recall that
$\mathfrak{D}$ denotes $n(d-1)d^n$. Then, we have the following
statements:
\begin{itemize}
  \item[i)] The arithmetic complexity of Algorithm
    \ref{algo:Weak-RRC-Hermite} lies in
    \[O\ {\widetilde{~}}\left (\binom{t+\mathfrak{D}}{t} \
   2^{3t}\ n^{2t+1} d^{2nt+n+2t+1} \right ).\]
\item[ii)] Algorithm \ref{algo:RRC-Hermite}, which is probabilistic,
computes a set of semi-algebraic descriptions solving Problem
\eqref{problem:rrc} within
\[O\ {\widetilde{~}}\left ( \binom{t+\mathfrak{D}}{t}\ 2^{3t}\
n^{2t+1} d^{3nt+2(n+t)+1} \right )\]
arithmetic operations in $\QQ$ in case of success.
\item[iii)] The semi-algebraic descriptions output by Algorithm
\ref{algo:RRC-Hermite} consist of polynomials in $\QQ[\by]$ of degree
bounded by $\mathfrak{D}$.
\end{itemize}
\end{reptheorem}
\begin{proof}
As Assumption \eqref{assumption:E} holds, we have that
$\bm{w}_{\infty} = 1$ and $\bm{w}_{\cH}$ is the square-free part of
$\det(\cH)$.

Therefore, after computing the parametric Hermite matrix $\cH$ and its
determinant, whose complexity is given by Proposition
\ref{prop:complexity-HM}, Algorithm \ref{algo:Weak-RRC-Hermite}
essentially consists of computing sample points of the connected
components of the algebraic set $\RR^t \setminus V(\det(\cH))$.

By Corollary \ref{corollary:degree-minor}, the degree of $\det(\cH)$
is bounded by $\mathfrak{D}$. Applying
Corollary~\ref{cor:sampling}, we obtain the following arithmetic
complexity for this computation of sample points
\[O\ {\widetilde{~}}\left( \binom{t+\mathfrak{D}}{t}\ 2^{3t}
\mathfrak{D}^{2t+1} \right ) \simeq O\ {\widetilde{~}}\left (
\binom{t+\mathfrak{D}}{t}\ 2^{3t}\ n^{2t+1} d^{2nt+n+2t+1}
\right ).\]
Also by Corollary~\ref{cor:sampling}, the finite subset of $\QQ^t$
output by {\sf SamplePoints} has cardinal bounded by
$2^t\mathfrak{D}^t$. Thus, evaluating the specializations of $\cH$ at
those points and their signatures costs in total $O \left ( 2^t
\mathfrak{D}^t \left ( \delta^2 \binom{2\lambda + t}{t} +
\delta^{\omega +1/2} \right ) \right)$ arithmetic operations in $\QQ$
using \cite[Algorithm 8.43]{BPR}.

Therefore, the complexity of {\sf SamplePoints} dominates the whole
complexity of the algorithm. We conclude that Algorithm
\ref{algo:Weak-RRC-Hermite} runs within
\[O\ {\widetilde{~}}\left ( \binom{t+\mathfrak{D}}{t}\ 2^{3t}\
n^{2t+1} d^{2nt+n+2t+1} \right )\]
arithmetic operations in $\QQ$.

For Algorithm \ref{algo:RRC-Hermite}, we start by choosing randomly a
matrix $A$ and compute the matrix $\cH_{A} = A^T \cdot \cH \cdot
A$. Then, we compute the leading principal minors
$M_1,\ldots,M_{\delta}$ of $\cH_{A}$. Using Proposition
\ref{prop:complexity-HM}, this step admits the arithmetic complexity
bound
\[O\ \widetilde{~} \left( \delta\ \binom{t+\mathfrak{D}}{t} \left (
d^{2n} \binom{t + 2\lambda}{t} + d^{\omega n}\right) \right ).\]

Next, Algorithm \ref{algo:RRC-Hermite} computes sample points for the
connected components of the semi-algebraic set defined by
$\wedge_{i=1}^{\delta}M_i \ne 0$. Since the degree of each $M_i$ is
bounded by $\mathfrak{D}$, Corollary~\ref{cor:sampling} gives the
arithmetic complexity
\[O\ {\widetilde{~}}\left(\binom{t+\mathfrak{D}}{t} \
d^{nt+n}\ 2^{3t}\ \mathfrak{D}^{2t+1} \right) \simeq O\
{\widetilde{~}}\left ( \binom{t+\mathfrak{D}}{t}\ 2^{3t}\
  n^{2t+1} d^{3nt+2(n+t)+1} \right ).\]
It returns a finite subset of $\QQ^t$ whose cardinal is bounded by
$\left ( 2\delta \mathfrak{D} \right) ^t $. The evaluation of the
leading principal minors' sign patterns at those points has the
arithmetic complexity lying in $O \left ( 2^t\delta^{t+1}
\mathfrak{D}^{2t} \right) \simeq O\left ( 2^t n^{2t} d^{3nt+n+2t}
\right)$.

Again, the complexity of {\sf SamplePoints} dominates the whole
complexity of Algorithm~\ref{algo:RRC-Hermite}. The proof of Theorem
\ref{thm:main} is then finished.
\end{proof}

\paragraph{Probability aspect} The main probabilistic source of
our algorithms~\ref{algo:Weak-RRC-Hermite} and \ref{algo:RRC-Hermite}
comes from the use of the geometric resolution \cite{GLS01} in the
computation of sample points per connected components described in
Section~\ref{section:sample-points}. Since the geometric resolution
depends on the specialization and lifting procedures, it makes use of
various random choices. As explained in \cite{GLS01}, the bad choices
are enclosed in strict algebraic subsets of certain affine spaces,
which implies that almost any random choice leads to a correct
computation. In general, even though one can check whether the points
output by geometric resolution are solutions of the input system, some
solutions can be missing. Thus, the geometric resolution is not Las
Vegas.

Besides, Algorithm~\ref{algo:RRC-Hermite} depends also on the choice
of the matrix $Q$. By Lemma~\ref{lemma:random-matrix}, any choice of
$Q$ from a prescribed dense Zariski open subset of ${\rm GL}(n,\CC)$
will work. As the purpose of choosing $Q$ is to ensure that none of
the leading principal minors of $Q^T\cdot \cH \cdot Q$ are identically
zero. One can check easily whether a good matrix $Q$ is found.

%% file: experiments.tex
\section{Practical implementation \& Experimental results}
\label{section:experiments}
\subsection{Remark on the implementation of
  Algorithm~\ref{algo:RRC-Hermite}}
\label{ssection:implementation}


Recall that Algorithm \ref{algo:RRC-Hermite} leads us to compute
sample points per connected components of the non-vanishing set of the
leading principal minors $(M_1,\ldots,M_{\delta})$. Comparing to
Algorithm~\ref{algo:Weak-RRC-Hermite} in which we only compute sample
points for $\RR^t \setminus V(M_{\delta})$, the complexity of
Algorithm~\ref{algo:RRC-Hermite} contains an extra factor of $d^{nt}$
due to the higher number of polynomials given as input to the
subroutine {\sf SamplePoints}. Even though the complexity bounds of
these two algorithms both lie in $d^{O(nt)}$, the extra factor
$d^{nt}$ mentioned above sometimes becomes the bottleneck of
Algorithm~\ref{algo:RRC-Hermite} for tackling practical problems.
Therefore, we introduce the following optimization in our
implementation of Algorithm~\ref{algo:RRC-Hermite}.

We start by following exactly the steps
(\ref{algo-line:1}-\ref{algo-line:2}) of
Algorithm~\ref{algo:RRC-Hermite} to obtain the leading principal
minors $(M_1,\ldots,M_{\delta})$ and the polynomial
$\bm{w}_{\infty}$. Then, by calling the subroutine {\sf SamplePoints}
on the input $M_{\delta} \ne 0 \wedge \bm{w}_{\infty} \ne 0$, we
compute a set of sample points (and their corresponding numbers of
real roots) $\{(\bma_1,r_1),\ldots,(\bma_{\ell},r_{\ell})\}$ that
solves the weak-version of Problem \eqref{problem:rrc}. We obtain from
this output all the possible numbers of real roots that the input
system can admit.

For each value $0 \leq r \leq \delta$, we define
\[\Phi_r = \{ \sigma = (\sigma_1,\ldots,\sigma_{\delta}) \in
\{-1,1\}^{\delta} \; | \;\text{ the sign variation of }\sigma\text{ is
}(\delta-r)/2 \}.\]
If $r \not\equiv \delta \pmod{2}$, $\Phi_r = \emptyset$.

For $\sigma \in \Phi_r$ and $\bma \in \RR^t\setminus
V(\bm{w}_{\infty})$ such that $\sign(M_i(\bma)) = \sigma_i$ for every
$1\leq i \leq \delta$, the signature of $\cH(\bma)$ is $r$. As a consequence,
for any $\bma$ in the semi-algebraic set defined by
\[(\bm{w}_{\infty} \ne 0) \wedge (\vee_{\sigma \in
    \Phi_r}(\wedge_{i=1}^{\delta} \sign(M_i) = \sigma_i)),\]
the system $\ff(\bma,.)$ has exactly $r$ distinct real solutions.

Therefore, $(\mathcal{S}_{r_i})_{1\leq i \leq \ell}$ is a collection
of semi-algebraic sets solving Problem~\eqref{problem:rrc}. Then, we
can simply return $\{(\Phi_{r_i},\bma_i, r_i)\; | \; 1\leq i \leq
\ell\}$ as the output of Algorithm~\ref{algo:RRC-Hermite} without any
further computation. Note that, by doing so, we may return sign conditions
which are not realizable.

We discuss now about the complexity aspect of the steps described
above. For $r \equiv \delta \pmod{2}$, the cardinal of $\Phi_r$ is
$\binom{\delta}{(\delta-r-2)/2}$. In theory, the total cardinal of all
the $\Phi_{r_i}$'s ($1\leq i\leq \ell$) can go up to $2^{\delta-1}$, which
is doubly exponential in the number of variables $n$. However, in the
instances that are actually tractable by the current state of the art,
$2^{\delta}$ is still smaller than $\delta^{3t}$. And when it is the case,
following this approach has better performance than computing the
sample points of the semi-algebraic set defined by
$\wedge_{i=1}^{\delta} M_i \ne 0$. Otherwise, when $2^{\delta}$
exceeds $\delta^{3t}$, we switch back to the computation of sample
points.

This implementation of Algorithm~\ref{algo:RRC-Hermite} does not
change the complexity bound given in Theorem~\ref{thm:main}.
\revi{\subsection{Implementation infrastructure}}

\revi{To implement our algorithm, we need three main ingredients: {\em (i)}
  Gr\"obner bases computations, in order to obtain monomial basis of quotient
  algebras that we use to compute our parametrized Hermite matrices, {\em (ii)}
  an implementation of an algorithm computing sample points connected components
of semi-algebraic sets, {\em (iii)} a computer algebra system to manipulate
polynomials and matrices.}

\revi{In our implementation, we use the \textsc{Maple} computer algebra system
  and its programming language to implement the overall algorithm. We use
  J.-Ch~Faug\`ere's {\sc FGb} library \cite{FGb}, implemented in C, for computing Gr\"obner
  bases. }

\revi{In order to compute sample points per connected components of
  semi-algebraic sets, we use the \textsc{RAGlib} \cite{RAG} (Real Algebraic
  Library) package which is implemented using the {\sc Maple} programming
  language and the {\sc FGb} library. The algorithm implemented therein is the
  one of \cite{FMRSa08} and its complexity remains to be established. Even if
  they share similar ingredients, it is not the same as the one of
  Section~\ref{section:sample-points} which provides the state-of-the-art
  complexity result for this problem. Hence, our implementation might not meet
  the best promised by complexity results. Still, we see in the experiments
  below that it already can tackle problems which are out of reach of the
  current software state-of-the-art.}

\subsection{Experiments}
This subsection provides numerical results of several algorithms related
to the real root classification. We report on the performance of each
algorithm for different test instances.

The computation is carried out on a computer of Intel(R) Xeon(R) CPU
E7-4820 2GHz and 1.5 TB of RAM. The timings are given in seconds (s.),
minutes (m.) and hours (h.). The symbol $\infty$ means that the
computation cannot finish within $120$ hours.

Throughout this subsection, the column {\sc hermite} reports on the
computational data of our algorithms based on parametric Hermite
matrices described in Section \ref{section:algorithm}. It uses the
notations below:
\begin{itemize}
  \item[-] {\sc mat}: the timing for computing a parametric Hermite
    matrix $\cH$.
   \item[-] {\sc det}: the runtime for computing the determinant of
$\cH$.
\item[-] {\sc min}: the timing for computing the leading principal
  minors of $\cH$ .
\item[-] {\sc sp}: the runtime for computing at least one points
per each connected component of the semi-algebraic set $\RR^t
\setminus \VV(\det(\cH))$.
\item[-] {\sc deg}: the highest degree among the leading principal
  minors of $\cH$.
\end{itemize}

\paragraph{Generic systems}
In this paragraph, we report on the results obtained with generic
inputs, i.e., randomly chosen dense polynomials
$(f_1,\ldots,f_n)\subset \QQ[y_1,\ldots,y_t][x_1,\ldots,x_n]$. The total
degrees of input polynomials are given as a list $d =
[\deg(f_1),\ldots,\deg(f_n)]$.

We first compare the algorithms using Hermite matrices (Section
\ref{section:algorithm}) with the folklore Sturm-based algorithm
sketched in the introduction for solving Problem
\eqref{problem:rrc}. The column {\sc sturm} of
Fig. \eqref{fig:generic-1} shows the experimental results of the
Sturm-based algorithm. It contains the following sub-columns:
\begin{itemize}
\item[-] {\sc elim}: the timing for computing the eliminating
  polynomial.
\item[-] {\sc sres}: the timing for computing the subresultant
  coefficients in the Sturm-based algorithm.
\item[-] {\sc sp-s}: the timing for computing sample points per
  connected components of the non-vanishing set of the last
  subresultant coefficient.
\item[-] {\sc deg-s}: the highest degree among the subresultant
  coefficients. 
\end{itemize}
We observe that the sum of {\sc mat-h} and {\sc min-h} is smaller than
the sum of {\sc elim} and {\sc sres}. Hence, obtaining the input for
the sample point computation in {\sc hermite} strategy is easier than
in {\sc sturm} strategy. We also remark that the degree {\sc deg-h} is
much smaller than {\sc deg-s}, that explains why the computation of
sample points using Hermite matrices is faster than using the
subresultant coefficients.

We conclude that the parametric Hermite matrix approach outperforms
the Sturm-based one both on the timings and the degree of polynomials
in the output formulas.

\begin{figure}[!ht]
\small
\centering
\begin{tabularx}{\textwidth}{|YY|YYYYY|YYYYY|}
\toprule
$t$ & $d$ & \multicolumn{5}{c|}{{\sc hermite}} &
\multicolumn{5}{c|}{{\sc sturm}} \\
& & {\sc mat} & {\sc min} & {\sc
sp} & {\bf total} & {\sc deg} & {\sc elim} & {\sc sres} & {\sc sp-s} &
 {\bf total} & {\sc deg-s} \\
\midrule
$2$ & $[2,2]$ & .07 s& .01 s& .3 s & .4 s & 8 & .01 s & .1 s & 2 s &
2.2 s & 12 \\
$2$ & $[3,2]$ & .1 s& .12 s & 4.8 s & 5 s & 18 & .05 s& .5 s& 15 s & 16
s & 30 \\
$2$ & {\footnotesize $[2,2,2]$} & .3 s & .3 s& 33 s & 34 s & 24 & .08 s& 2 s& 8
m & 8 m & 56 \\
$2$ & $[3,3]$ & .3 s & .8 s& 3 m & 3 m & 36 & .1 s& 3 s& 20 m & 20 m &
72 \\
\midrule
$3$ & $[2,2]$ & .1 s& .02 s& 26 s & 27 s & 8 & .07 s & .1 s & 40
s & 40 s & 12 \\
$3$ & $[3,2]$ & .2 s& .2 s& 3 h & 3 h & 18 & .1 s & 1 s & $\infty$
& $\infty$ & 30 \\
$3$ & {\footnotesize $[2,2,2]$} & .5 s& 7 s & 32 h & 32 h & 24 & .15 s & 10 m &
$\infty$ & $\infty$ & 56 \\
$3$ & $[4,2]$ & .6 s & 12 s & 90 h & 90 h & 32 & .2 s &
 12 m & $\infty$ & $\infty$ & 56 \\
$3$ & $[3,3]$ & 1 s& 27 s& $\infty$ & $\infty$ & 36 & .2 s & 15 m &
$\infty$ & $\infty$ & 72 \\
\bottomrule
\end{tabularx}
\caption{Generic random dense systems}
\label{fig:generic-1}
\end{figure}

In Fig. \eqref{fig:generic-2}, we compare our algorithm using
parametric Hermite matrices with two Maple packages for solving
parametric polynomial systems: {\sc RootFinding[Parametric]}
\cite{GeJeMo10} and {\sc RegularChains[ParametricSystemTools]}
\cite{YangHx01}. The new notations used in Fig. \eqref{fig:generic-2}
are explained below.

\begin{itemize}
  \item The column {\sc rf} stands for the
\textsc{RootFinding[Parametric]} package. To solve a parametric
polynomial systems, it consists of computing a discriminant variety
$\mathcal{D}$ and then computing an open CAD of $\RR^t\setminus
\mathcal{D}$. \emph{This package does not return explicit semi-algebraic
formulas but an encoding based on the real roots of some polynomials.}

This column contains:
\begin{itemize}
  \item[-] {\sc dv} : the runtime of the command
\textsc{DiscriminantVariety} that computes a set of polynomials
defining a discriminant variety $\mathcal{D}$ associated to the input
system.
  \item[-] {\sc cad} : the runtime of the command
\textsc{CellDecomposition} that outputs semi-algebraic formulas by
computing an open CAD for the semi-algebraic set $\RR^t\setminus
\mathcal{D}$.
\end{itemize}
\item[$\bullet$] The column {\sc rc} stands for the {\sc
RegularChains[ParametricSystemTools]} package of Maple. The algorithms
implemented in this package is given in \cite{YangHx01}. It also
contains two sub-columns:
\begin{itemize}
\item[-] {\sc bp} : the runtime of the command {\sc BorderPolynomial}
  that returns a set of polynomials.
\item[-] {\sc rrc} : the runtime of the command {\sc
RealRootClassification}. We call this command with the option
\texttt{output=`samples'} to compute at least one point per connected
component of the complementary of the real algebraic set defined by
border polynomials.
\end{itemize}
\end{itemize}

Note that, in a strategy for solving the weak-version of Problem
\eqref{problem:rrc}, {\sc DiscriminantVariety} and {\sc
BorderPolynomial} can be completely replaced by parametric Hermite
matrices.

On generic systems, the determinant of our parametric Hermite matrix
coincides with the output of {\sc DiscriminantVariety}, which we
denote by $\bm{w}$. Whereas, because of the elimination {\sc
BorderPolynomial} returns several polynomials, one of them is
$\bm{w}$.

In Fig. \eqref{fig:generic-2}, the timings for computing a parametric
Hermite matrix is negligible. Comparing the columns {\sc det}, {\sc
dv} and {\sc bp}, we remark that the time taken to obtain $\bm{w}$
through the determinant of parametric Hermite matrices is much smaller
than using {\sc DiscriminantVariety} or {\sc BorderPolynomial}.

For computing the polynomial $\bm{w}$, using parametric Hermite
matrices allows us to reach the instances that are out of reach of
\textsc{DiscriminantVariety}, for example, the instances $\{t=3,\; d =
[2,2,2]\}$, $\{t=3\; d = [4,2]\}$, $\{t=3,\; d=[3,3]\}$ and $\{t =
4,\; d = [2,2]\}$ in Fig. \eqref{fig:generic-2} below. Moreover, we
succeed to compute the semi-algebraic formulas for $\{t=3,\; d =
[2,2,2]\}$, $\{t=3\; d = [4,2]\}$ and $\{t = 4,\; d = [2,2]\}$. Using
the implementation in Subsection~\ref{ssection:implementation}, we
obtain the semi-algebraic formulas of degrees bounded by
$\deg(\bm{w})$.

Therefore, for these generic systems, our algorithm based on
parametric Hermite matrices outperforms \textsc{DiscriminantVariety}
and \textsc{BorderPolynomial} for obtaining a polynomial that defines
the boundary of semi-algebraic sets over which the number of real
solutions are invariant. Moreover, using the minors of parametric
Hermite matrices, we can compute semi-algebraic formulas of problems
that are out of reach of {\sc CellDecomposition} and {\sc
  RealRootClassification}. 

\begin{figure}[!ht]
\small
\centering
\begin{tabularx}{\textwidth}{|YY|YYYYY|YYY|YYY|}
\toprule
$t$ & $d$ & \multicolumn{5}{c|}{{\sc hermite}} &
\multicolumn{3}{c|}{{\sc rf}} & \multicolumn{3}{c|}{\sc rc} \\
& & {\sc mat} & {\sc det} & {\sc sp} & {\bf total} & {\sc
deg} & {\sc dv} & {\sc cad} & {\bf total} & {\sc bp} & {\sc rrc} &
{\bf total} \\
\midrule
$2$ & $[2,2]$ & .07 s& .01 s& .3 s& .4 s & 8 & .1 s& .3 s& .4 s &  .1
s& 1 s & 1.1 s \\
$2$ & $[3,2]$ & .1 s& .2 s& 4.8 s& 5 s & 18 & 1 m & 5 s& 1 m & .3 s&
12 s & 12 s \\
$2$ & {\footnotesize $[2,2,2]$} & .3 s& .3 s& 33 s& 34 s & 24 & 17m & 32 s& 17m & 23
s& 2 m & 2 m \\
$2$ & $[3,3]$ & .3 s & .8 s& 3 m & 3 m & 36 & 2 h & 4 m & 2 h & 8 s &
4 m & 4 m \\
\midrule
$3$ & $[2,2]$ & .1 s& .02 s& 26 s& 27 s & 8 & 1 s & 35 s & 36 s & .2
s& 12m & 12m \\
$3$ & $[3,2]$ & .2 s& .2 s& 3 h & 3 h & 18 & 2 h & 84
h & 86 h & 3 s & 37 h & 37 h \\
$3$ & {\footnotesize $[2,2,2]$} & .5 s& 7 s & 32 h & 32 h & 24 & $\infty$ &
$\infty$ & $\infty$ & 20m & $\infty$ & $\infty$ \\
$3$ & $[4,2]$ & .6 s & 12 s & 90 h & 90 h & 32 & $\infty$ &
$\infty$ & $\infty$ & 12m & $\infty$ & $\infty$ \\
$3$ & $[3,3]$ & .7 s& 27 s& $\infty$ & $\infty$ & 36 & $\infty$ &
$\infty$ & $\infty$ & 15m & $\infty$ & $\infty$ \\
\midrule
$4$ & $[2,2]$ & .2 s& .1 s& 8 m & 8 m & 8 & 4 s &
$\infty$ & $\infty$ & 1 s & $\infty$ & $\infty$ \\
\bottomrule
\end{tabularx}
\caption{Generic random dense systems}
\label{fig:generic-2}
\end{figure}


In what follows, we consider the systems coming from some applications
as test instances. These examples allow us to observe the behavior of
our algorithms on non-generic systems.

\paragraph{Kuramoto model} This application is introduced in
\cite{Kura75}, which is a dynamical system used to model
synchronization among some given coupled oscillators. Here we consider
only the model constituted by $4$ oscillators. The maximum number of
real solutions of steady-state equations of this model was an open
problem before it is solved in \cite{Harris20} using numerical
homotopy continuation methods. However, to the best of our knowledge,
there is no exact algorithm that is able to solve this problem. We
present in what follows the first solution using symbolic
computation. Moreover, our algorithm can return the semi-algebraic
formulas defining the regions over which the number of real solutions
is invariant.

As explained in \cite{Harris20}, we consider the system $\ff$ of the
following equations
\[\left\{\begin{array}{ll}
           y_i-\sum_{j=1}^4(s_ic_j-s_jc_i) & = 0 \\
           s_i^2+c_i^2 & = 1
         \end{array}\right.\text{ for }1\leq i \leq 3,\]
where $(s_1,s_2,s_3)$ and $(c_1,c_2,c_3)$ are variables and
$(y_1,y_2,y_3)$ are parameters. We are asked to compute the maximum
number of real solutions of $\ff(\bma,.)$ when $\bma$ varies over
$\RR^3$. This leads us to solve the weak version of Problem
\eqref{problem:rrc} for this parametric system.

We first construct the parametric Hermite matrix $\cH$ associated to
this system. This matrix is of size $14\times 14$. The polynomial
$\bm{w}_{\infty}$ has the factors $y_1+y_2$, $y_2+y_3$, $y_3+y_1$ and
$y_1+y_2+y_3$. The polynomial $\bm{w}_{\cH}$ has degree $48$
(c.f. \cite{Harris20}). We denote by $\bm{w}$ the polynomial
$\bm{w}_{\infty} \cdot \bm{w}_{\cH}$.

Note that the polynomial system has real roots only if $|y_i|\leq 3$
(c.f. \cite{Harris20}). So we only need to consider the compact
connected components of $\RR^3\setminus \VV(\bm{w})$. Since the
polynomial $\bm{w}$ is invariant under any permutation acting on
$(y_1,y_2,y_3)$, we exploit this symmetry to accelerate the
computation of sample points.

Following the critical point method, we compute the critical points of
the map $(y_1,y_2,y_3) \mapsto y_1+y_2+y_3$ restricted to
$\RR^3\setminus \VV(\bm{w})$; this map is also symmetric. We apply the
change of variables
\[(y_1,y_2,y_3) \mapsto (e_1,e_2,e_3),\]
where $e_1=y_1+y_2+y_3$, $e_2=y_1y_2+y_2y_3+y_3y_1$ and
$e_3=y_1y_2y_3$ are elementary symmetric polynomials of
$(y_1,y_2,y_3)$. This change of variables reduces the number of
distinct solutions of zero-dimensional systems involved in the
computation and, therefore, reduces the computation time.

From the sample points obtained by this computation, we derive the
possible number of real solutions and conclude that the system $\ff$
has at most $10$ distinct real solutions when $(y_1,y_2,y_3)$ varies
over $\RR^3 \setminus V(\bm{w})$. This agrees with the result given
in \cite{Harris20}. We show below a list of parameter values such that
the system has respectively $2$, $4$, $6$, $8$ and $10$ distinct real
solutions.
\begin{center}
\begin{tabular}{ c | c }
Number of solutions & $(y_1,y_2,y_3)$ \\
\midrule
2 solutions & $[-2, -0.03, 0.22]$ \\
4 solutions & $[1, -0.09, 0.16]$ \\ 
6 solutions & $[0, -0.7, -0.48]$ \\
8 solutions & $[0.08,-0.03, 0.22]$ \\
10 solutions & $\left[ \frac{274945023031}{2199023255552},
               \frac{-68723139707}{549755813888},
               \frac{-549808278091}{4398046511104} \right]$
\end{tabular}
\end{center}

Fig. \eqref{fig:kuramoto} reports on the timings for computing the
parametric Hermite matrix ({\sc mat}), for computing its determinant
({\sc det}) and for computing the sample points ({\sc sp}). We
stop both of the commands {\sc DiscriminantVariety} and {\sc
BorderPolynomial} after $240$ hours without obtaining the polynomial
$\bm{w}$.

\begin{figure}[!ht]
  \small
  \centering
  \begin{tabularx}{\textwidth}{|YYYY|Y|Y|}
    \toprule
\multicolumn{4}{|c|}{\sc hermite} & {\sc dv} & {\sc bp} \\
{\sc mat} & {\sc det} & {\sc sp} & {\bf total} & & \\
  \hline
  $2$ m & $1$ h & $85$ h & 86 h & $\infty$ & $\infty$ \\
\bottomrule
\end{tabularx}
\caption{Kuramoto model for $4$ oscillators}
  \label{fig:kuramoto}
\end{figure}

\paragraph{Static output feedback}
The second non-generic example comes from the problem of static output
feedback \cite{Henrion08}. Given the matrices $A \in \RR^{\ell \times
\ell}$, $B\in \RR^{\ell \times 2}$, $C \in \RR^{1\times \ell}$ and a
parameter vector $P = \begin{bmatrix} y_1 \\ y_2\end{bmatrix} \in
\RR^2$, the characteristic polynomial of $A+BPC$ writes
\[f(s,\by) = \det(sI_l - A - BKC) = f_0(s)+y_1f_1(s)+y_2f_2(s),\]
where $s$ is a complex variable.

We want to find a matrix $P$ such that all the roots of $f(s,\by)$
must lie in the open left half-plane. By substituting $s$ by
$x_1+ix_2$, we obtain the following system of real variables
$(x_1,x_2)$ and parameters $(y_1,y_2)$:
\[\left\{\begin{array}{ll} \Re(f(x_1+ix_2,\by)) & = 0 \\
\Im(f(x_1+ix_2,\by)) & = 0 \\ x_1 & < 0
         \end{array}\right.\]
Note that the total degree of these equations equals $\ell$.
     
We are now interested in solving the weak-version of Problem
\eqref{problem:rrc} on the system $\Re(f) = \Im(f)=0$. We observe that
this system satisfies Assumptions \eqref{assumption:A} and
\eqref{assumption:C}. Let $\cH$ be the parametric Hermite matrix $\cH$
of this system with respect to the usual basis we consider in this
paper. This matrix $\cH$ behaves very differently from generic
systems.

Computing the determinant of $\cH$ (which is an element of $\QQ[\by]$)
and taking its square-free part allows us to obtain the same output
$\bm{w}$ as {\sc DiscriminantVariety}. However, this direct
approach appears to be very inefficient as the determinant appears as
a large power of the output polynomial.

For example, for a value $\ell$, we observe that the system consists
of two polynomials of degree $\ell$. The determinant of $\cH$ appears
as $\bm{w}^{2\ell}$, where $\bm{w}$ has degree $2(\ell-1)$. The bound
we establish on the degree of this determinant is $2(\ell-1)\ell^2$,
which is much larger than what happens in this case. Therefore, we
need to introduce the optimization below to adapt our implementation
of Algorithm~\ref{algo:Weak-RRC-Hermite} to this problem.

We observe that, on these examples, the polynomial $\bm{w}$ can be
extracted from a smaller minor instead of computing the determinant
$\cH$. To identify such a minor, we reduce $\cH$ to a matrix whose
entries are univariate polynomials with coefficients lying in a finite
field $\mathbb{Z}/p \mathbb{Z}$ as follow. 

Let $u$ be a new variable. We substitute each $y_i$ by random linear
forms in $\QQ[u]$ in $\cH$ and then compute $\cH \bmod p$. Then,
the matrix $\cH$ is turned into a matrix $\cH_u$ whose entries are
elements of $\mathbb{Z}/p\mathbb{Z}[u]$. The computation of the leading
principal minors of $\cH_u$ is much easier than the one of $\cH$ since
it involves only univariate polynomials and does not suffer from the
growth of bit-sizes as for the rational numbers.

Next, we compute the sequence of the leading principal minors of
$\cH_u$ in decreasing order, starting from the determinant. Once we
obtain a minor, of some size $r$, that is not divisible by
$\overline{\bm{w}}_u$, we stop and take the index $r+1$.  Then, we
compute the square-free part of the $(r+1)\times (r+1)$ leading
principal minor of $\cH$, which can be done through
evaluation-interpolation method. This yields a Monte Carlo
implementation that depends on the choice of the random linear forms
in $\QQ[u]$ and the finite field to compute the polynomial $\bm{w}$.

In Fig. \eqref{fig:static-output-feedback}, we report on some
computational data for the static output feedback problem. Here we
choose the prime $p$ to be $65521$ so that the elements of the finite
field $\mathbb{Z}/p\mathbb{Z}$
can be represented by a machine word of $32$ bits. We consider
different values of $\ell$ and the matrices $A,B,C$ are chosen
randomly. On these examples, our algorithm returns the same output as
the one of \textsc{DisciminantVariety}. Whereas, {\sc
BorderPolynomial} ({\sc bp}) returns a list of polynomials which
contains our output and other polynomials of higher degree.

The timings of our algorithm are given by the two following columns:
\begin{itemize}
\item The column {\sc mat} shows the timings for computing
  parametric Hermite matrices $\cH$.
\item The column {\sc comp-w} shows the timings for computing the
  polynomials $\bm{w}$ from $\cH$ using the strategy described as above.
\end{itemize}
We observe that our algorithm ({\sc mat} + {\sc comp-w})
wins some constant factor comparing to {\sc DiscriminantVariety} ({\sc
dv}). On the other hand, {\sc BorderPolynomial} ({\sc bp}) performs less
efficiently than the other two algorithms in these examples.

Since the degrees of the polynomials $\bm{w}$ here (given as {\sc
deg-w}) are small comparing with the bounds in the generic
case. Hence, unlike the generic cases, the computation of the sample
points in these problems is negligible as being reported in the column
{\sc sp}.

\begin{figure}[!ht]
  \small
  \centering
  \begin{tabularx}{\textwidth}{|Y | Y Y Y | Y | Y | Y | Y|}
    \toprule
  $\ell$ & \multicolumn{3}{c|}{\sc hermite} & {\sc dv} & {\sc bp} &
{\sc sp} & {\sc deg-w}\\
  & {\sc mat} & {\sc comp-w} & {\bf total} & & & & \\
  \midrule
  $5$ & $2$ s & $1$ s & 3 s & $30$ s & $1.5$ m & $.2$ s & $8$ \\
  $6$ & $12$ s & $5$ s & 17 s & $90$ s & $30$ m & $.4$ s & $10$ \\
  $7$ & $1$ m & $6$ m & 7 m & $16$ m & $4$ h & $1$ s & $12$ \\
  $8$ & $4$ m & $50$ m & 1 h & $1.5$ h & $34$ h & $3$ s & $14$ \\
  \bottomrule
\end{tabularx}
\caption{Static output feedback}
\label{fig:static-output-feedback}
\end{figure}


%% file: mainRRC.bbl
\begin{thebibliography}{10}

\bibitem{Alman21}
{\sc Alman, J., and Williams, V.~V.}
\newblock A refined laser method and faster matrix multiplication.
\newblock In {\em Proceedings of the Thirty-Second Annual ACM-SIAM Symposium on
  Discrete Algorithms\/} (USA, 2021), SODA '21, Society for Industrial and
  Applied Mathematics, p.~522–539.

\bibitem{Bar-thesis}
{\sc Bardet, M.}
\newblock {\em {{\'E}tude des syst{\`e}mes alg{\'e}briques
  surd{\'e}termin{\'e}s. Applications aux codes correcteurs et {\`a} la
  cryptographie}}.
\newblock Theses, {Universit{\'e} Pierre et Marie Curie - Paris VI}, Dec. 2004.

\bibitem{BFS14}
{\sc Bardet, M., Faug{\`e}re, J.-C., and Salvy, B.}
\newblock On the complexity of the {F}5 {G}r{\"o}bner basis algorithm.
\newblock {\em Journal of Symbolic Computation 70\/} (2015), 49--70.

\bibitem{BPR}
{\sc Basu, S., Pollack, R., and Roy, M.-F.}
\newblock {\em Algorithms in Real Algebraic Geometry (Algorithms and
  Computation in Mathematics)}.
\newblock Springer-Verlag, Berlin, Heidelberg, 2006.

\bibitem{BaSt87}
{\sc Bayer, D., and Stillman, M.}
\newblock A theorem on refining division orders by the reverse lexicographic
  order.
\newblock {\em Duke Math. J. 55}, 2 (06 1987), 321--328.

\bibitem{BFJSV16}
{\sc Bonnard, B., Faug{\`e}re, J.-C., Jacquemard, A., Safey El~Din, M., and
  Verron, T.}
\newblock Determinantal sets, singularities and application to optimal control
  in medical imagery.
\newblock In {\em Proceedings of the ACM on International Symposium on Symbolic
  and Algebraic Computation\/} (2016), pp.~103--110.

\bibitem{Dav07}
{\sc Brown, C.~W., and Davenport, J.~H.}
\newblock The complexity of quantifier elimination and cylindrical algebraic
  decomposition.
\newblock In {\em Proceedings of the 2007 International Symposium on Symbolic
  and Algebraic Computation\/} (New York, NY, USA, 2007), ISSAC ’07,
  Association for Computing Machinery, p.~54–60.

\bibitem{CaKaYa89}
{\sc Canny, J.~F., Kaltofen, E., and Yagati, L.}
\newblock Solving systems of nonlinear polynomial equations faster.
\newblock In {\em Proceedings of the ACM-SIGSAM 1989 International Symposium on
  Symbolic and Algebraic Computation\/} (New York, NY, USA, 1989), ISSAC '89,
  Association for Computing Machinery, p.~121–128.

\bibitem{Collins76}
{\sc Collins, G.~E.}
\newblock Quantifier elimination for real closed fields by cylindrical
  algebraic decomposition: a synopsis.
\newblock {\em {ACM} {SIGSAM} Bulletin 10}, 1 (1976), 10--12.

\bibitem{Co02}
{\sc Corvez, S., and Rouillier, F.}
\newblock Using computer algebra tools to classify serial manipulators.
\newblock In {\em International Workshop on Automated Deduction in Geometry\/}
  (2002), Springer, pp.~31--43.

\bibitem{CoSh92}
{\sc Coste, M., and Shiota, M.}
\newblock Thom's first isotopy lemma: a semialgebraic version, with uniform
  bounds(real singularities and real algebraic geometry).
\newblock {\em RIMS Kokyuroku 815\/} (dec 1992), 176--189.

\bibitem{CLO}
{\sc Cox, D.~A., Little, J., and O’Shea, D.}
\newblock {\em Ideals, Varieties, and Algorithms: An Introduction to
  Computational Algebraic Geometry and Commutative Algebra, 3/e (Undergraduate
  Texts in Mathematics)}.
\newblock Springer-Verlag, Berlin, Heidelberg, 2007.

\bibitem{DS04}
{\sc Dahan, X., and Schost, {\'{E}}.}
\newblock Sharp estimates for triangular sets.
\newblock In {\em Symbolic and Algebraic Computation, International Symposium
  {ISSAC} 2004, Santander, Spain, July 4-7, 2004, Proceedings\/} (2004),
  J.~Gutierrez, Ed., {ACM}, pp.~103--110.

\bibitem{Dav88}
{\sc Davenport, J.~H., and Heintz, J.}
\newblock Real quantifier elimination is doubly exponential.
\newblock {\em J. Symb. Comput. 5}, 1–2 (Feb. 1988), 29–35.

\bibitem{EGS20}
{\sc Elliott, J., Giesbrecht, M., and Schost, {\'{E}}.}
\newblock On the bit complexity of finding points in connected components of a
  smooth real hypersurface.
\newblock In {\em {ISSAC} '20: International Symposium on Symbolic and
  Algebraic Computation, Kalamata, Greece, July 20-23, 2020\/} (2020), I.~Z.
  Emiris and L.~Zhi, Eds., {ACM}, pp.~170--177.

\bibitem{Fa13}
{\sc Faug{\`{e}}re, J., Gaudry, P., Huot, L., and Renault, G.}
\newblock Polynomial systems solving by fast linear algebra.
\newblock {\em CoRR abs/1304.6039\/} (2013).

\bibitem{F4}
{\sc Faugere, J.-C.}
\newblock A new efficient algorithm for computing {G}r{\"o}bner bases ({F}4).
\newblock {\em Journal of pure and applied algebra 139}, 1-3 (1999), 61--88.

\bibitem{F5}
{\sc Faug{\`e}re, J.~C.}
\newblock A new efficient algorithm for computing {G}r{\"o}bner bases without
  reduction to zero ({F}5).
\newblock In {\em Proceedings of the 2002 international symposium on Symbolic
  and algebraic computation\/} (2002), pp.~75--83.

\bibitem{FMRSa08}
{\sc Faug{\`e}re, J.-C., Moroz, G., Rouillier, F., and Safey El~Din, M.}
\newblock Classification of the perspective-three-point problem, discriminant
  variety and real solving polynomial systems of inequalities.
\newblock In {\em Proceedings of the twenty-first international symposium on
  Symbolic and algebraic computation\/} (2008), pp.~79--86.

\bibitem{Spa14}
{\sc Faug{\`e}re, J.-C., Safey El~Din, M., and Spaenlehauer, P.-J.}
\newblock On the complexity of the generalized minrank problem.
\newblock {\em Journal of Symbolic Computation 55\/} (2013), 30--58.

\bibitem{FGb}
{\sc Faugère, J.-C.}
\newblock {FGb: A Library for Computing Gröbner Bases}.
\newblock In {\em {Mathematical Software - ICMS 2010}\/} (Berlin, Heidelberg,
  September 2010), K.~Fukuda, J.~Hoeven, M.~Joswig, and N.~Takayama, Eds.,
  vol.~6327 of {\em Lecture Notes in Computer Science}, Springer Berlin /
  Heidelberg, pp.~84--87.

\bibitem{GeJeMo10}
{\sc Gerhard, J., Jeffrey, D.~J., and Moroz, G.}
\newblock A package for solving parametric polynomial systems.
\newblock {\em ACM Commun. Comput. Algebra 43}, 3/4 (June 2010), 61–72.

\bibitem{GhysRa16}
{\sc Ghys, {\'E}., and Ranicki, A.}
\newblock {Signatures in algebra, topology and dynamics}.
\newblock {\em {Ensaios Matem{\'a}ticos} 30\/} (2016), 1 -- 173.

\bibitem{GianniM87}
{\sc Gianni, P.~M., and Teo~Mora, T.}
\newblock Algebraic solution of systems of polynomial equations using
  {G}r\"oebner bases.
\newblock In {\em Applied Algebra, Algebraic Algorithms and Error-Correcting
  Codes, 5th International Conference, AAECC-5, Menorca, Spain, June 15-19,
  1987, Proceedings\/} (1987), pp.~247--257.

\bibitem{GiustiHMP95}
{\sc Giusti, M., Heintz, J., Morais, J.~E., and Pardo, L.~M.}
\newblock When polynomial equation systems can be "solved" fast?
\newblock In {\em Applied Algebra, Algebraic Algorithms and Error-Correcting
  Codes, 11th International Symposium, AAECC-11, Paris, France, July 17-22,
  1995, Proceedings\/} (1995), pp.~205--231.

\bibitem{GLS01}
{\sc Giusti, M., Lecerf, G., and Salvy, B.}
\newblock A gr{\"o}bner free alternative for polynomial system solving.
\newblock {\em Journal of complexity 17}, 1 (2001), 154--211.

\bibitem{Hardt80}
{\sc Hardt, R.~M.}
\newblock Semi-algebraic local-triviality in semi-algebraic mappings.
\newblock {\em American Journal of Mathematics 102}, 2 (1980), 291--302.

\bibitem{Harris20}
{\sc Harris, K., Hauenstein, J.~D., and Szanto, A.}
\newblock Smooth points on semi-algebraic sets, 2020.

\bibitem{Henrion08}
{\sc Henrion, D., and Sebek, M.}
\newblock Plane geometry and convexity of polynomial stability regions.
\newblock In {\em Proceedings of the Twenty-First International Symposium on
  Symbolic and Algebraic Computation\/} (New York, NY, USA, 2008), ISSAC ’08,
  Association for Computing Machinery, p.~111–116.

\bibitem{Hermite56}
{\sc Hermite, C.}
\newblock Sur le nombre des racines d'une {\'e}quation alg{\'e}brique comprises
  entre des limites donn{\'e}es. extrait d'une lettre {\'a} m. borchardt.
\newblock {\em J. Reine Angew. Math. 52\/} (1856), 39--51.

\bibitem{Jacobi57}
{\sc Jacobi, C.~G.}
\newblock Uber eine elementare transformation eins in bezug auf jedes von zwei
  variablen-systemen linearen und homogenen ausdrucks.
\newblock {\em Journal fur die reine und angewandte Mathematik 53.\/} (1857),
  265 -- 270.

\bibitem{Kal97}
{\sc Kalkbrener, M.}
\newblock On the stability of gr{\"o}bner bases under specializations.
\newblock {\em Journal of Symbolic Computation 24}, 1 (1997), 51--58.

\bibitem{Kr82}
{\sc Kronecker, L.}
\newblock Grundz\"uge einer arithmetischen theorie der algebraischen gr\"ossen.
\newblock {\em Journal f\"ur die reine und angewandte Mathematik 92\/} (1882),
  1--122.

\bibitem{Kura75}
{\sc Kuramoto, Y.}
\newblock Self-entrainment of a population of coupled non-linear oscillators.
\newblock In {\em International Symposium on Mathematical Problems in
  Theoretical Physics\/} (Berlin, Heidelberg, 1975), H.~Araki, Ed., Springer
  Berlin Heidelberg, pp.~420--422.

\bibitem{LaRo07}
{\sc Lazard, D., and Rouillier, F.}
\newblock Solving parametric polynomial systems.
\newblock {\em Journal of Symbolic Computation 42}, 6 (2007), 636--667.

\bibitem{Moreno03}
{\sc Moreno-Soc{\i}as, G.}
\newblock Degrevlex gr{\"o}bner bases of generic complete intersections.
\newblock {\em Journal of Pure and Applied Algebra 180}, 3 (2003), 263 -- 283.

\bibitem{Pardue10}
{\sc Pardue, K.}
\newblock Generic sequences of polynomials.
\newblock {\em Journal of Algebra 324}, 4 (2010), 579 -- 590.

\bibitem{PRS93}
{\sc Pedersen, P., Roy, M.-F., and Szpirglas, A.}
\newblock Counting real zeros in the multivariate case.
\newblock In {\em Computational Algebraic Geometry\/} (Boston, MA, 1993),
  F.~Eyssette and A.~Galligo, Eds., Birkh{\"a}user Boston, pp.~203--224.

\bibitem{Ro99}
{\sc Rouillier, F.}
\newblock Solving zero-dimensional systems through the rational univariate
  representation.
\newblock {\em Appl. Algebra Eng. Commun. Comput. 9}, 5 (1999), 433--461.

\bibitem{RAG}
{\sc Safey El~Din, M.}
\newblock Real alebraic geometry library, raglib (version 3.4), 2017.

\bibitem{SaSc03}
{\sc {Safey El Din}, M., and Schost, E.}
\newblock Polar varieties and computation of one point in each connected
  component of a smooth real algebraic set.
\newblock In {\em Proc. of the 2003 Int. Symp. on Symb. and Alg. Comp.\/} (NY,
  USA, 2003), ISSAC ’03, ACM, p.~224–231.

\bibitem{SaSc17}
{\sc {Safey El Din}, M., and Schost, {\'{E}}.}
\newblock A nearly optimal algorithm for deciding connectivity queries in
  smooth and bounded real algebraic sets.
\newblock {\em J. ACM 63}, 6 (Jan. 2017), 48:1--48:37.

\bibitem{SaSc18}
{\sc {Safey El Din}, M., and Schost, E.}
\newblock Bit complexity for multi-homogeneous polynomial system
  solving—application to polynomial minimization.
\newblock {\em Journal of Symbolic Computation 87\/} (2018), 176 -- 206.

\bibitem{Sc03}
{\sc Schost, {\'E}.}
\newblock Computing parametric geometric resolutions.
\newblock {\em Applicable Algebra in Engineering, Communication and Computing
  13}, 5 (2003), 349--393.

\bibitem{Shafa}
{\sc Shafarevich, I.~R.}
\newblock {\em Basic Algebraic Geometry 1: Varieties in Projective Space}.
\newblock Springer Berlin Heidelberg, Berlin, Heidelberg, 2013.

\bibitem{Syl1852}
{\sc Sylvester, J.~J.}
\newblock A demonstration of the theorem that every homogeneous quadratic
  polynomial is reducible by real orthogonal substitution to the form of a sum
  of positive and negative squares.
\newblock {\em Philosophical Magazine IV.\/} (1852), 138 -- 142.

\bibitem{Verron16}
{\sc Verron, T.}
\newblock {\em {Regularisation of Gr{\"o}bner basis computations for weighted
  and determinantal systems, and application to medical imagery}}.
\newblock Theses, {Universit{\'e} Pierre et Marie Curie - Paris VI}, Sept.
  2016.

\bibitem{YangHx01}
{\sc Yang, L., Hou, X., and Xia, B.}
\newblock A complete algorithm for automated discovering of a class of
  inequality-type theorems.
\newblock {\em Science in China Series F Information Sciences 44}, 1 (2001),
  33--49.

\bibitem{YangXia05}
{\sc Yang, L., and Xia, B.}
\newblock Real solution classification for parametric semi-algebraic systems.
\newblock In {\em Algorithmic Algebra and Logic. Proceedings of the {A3L} 2005,
  April 3-6, Passau, Germany; Conference in Honor of the 60th Birthday of
  Volker Weispfenning\/} (2005), A.~Dolzmann, A.~Seidl, and T.~Sturm, Eds.,
  pp.~281--289.

\bibitem{Yang00}
{\sc Yang, L., and Zeng, Z.}
\newblock Equi-cevaline points on triangles.
\newblock In {\em Computer Mathematics: Proceedings of the Fourth Asian
  Symposium (ASCM 2000)\/} (2000), World Scientific Publishing Company
  Incorporated, p.~130.

\bibitem{YaZhe05}
{\sc Yang, L., and Zeng, Z.}
\newblock An open problem on metric invariants of tetrahedra.
\newblock In {\em Proceedings of the 2005 International Symposium on Symbolic
  and Algebraic Computation\/} (New York, NY, USA, 2005), ISSAC ’05,
  Association for Computing Machinery, p.~362–364.

\end{thebibliography}
